\documentclass[12pt,paper]{iopart}
\usepackage{amssymb,graphicx,graphics,rotating,bm}	
\usepackage{subcaption}
\usepackage{xcolor}
\usepackage{float}
\usepackage{dcolumn}
\usepackage{bm}
\usepackage{hyperref,color,soul,mathbbol}
\eqnobysec 

\newcommand{\ud}{\mathrm{d}}
\newcommand{\ic}{\mathrm{i}}

\newcommand{\zZ}{\mathbb{Z}}
\newcommand{\qQ}{\mathbb{Q}}

\newcommand{\arcsinh}{\mathrm{arcsinh}} 
\newcommand{\one}{\mathbb{I}}
\newcommand{\D}[1]{\displaystyle}

\newlength{\subfigwidth}
\begin{document} 
\title{Classical transport in a maximally chaotic chain}
\author{William~Alderson and R\'emy~Dubertrand}
\address{Department of Mathematics, Physics and Electrical Engineering, \\Northumbria University, NE1 8ST Newcastle upon Tyne, United Kingdom}
\author{Akira~Shudo}
\address{Department of Physics, Faculty of Science, Tokyo Metropolitan University, Tokyo 192-0397}
\ead{remy.dubertrand@northumbria.ac.uk}

\date{\today}%
\begin{abstract} 
A model for a lattice of coupled cat maps has been recently introduced. This new and specific choice of the coupling makes the description especially easy and nontrivial quantities as Lyapunov exponents determined exactly. We studied the ergodic property of the dynamics along such a chain for a local perturbation. While the perturbation spreads across a front growing ballistically, the position and momentum profiles show large fluctuations due to chaos leading to diffusive transport in the phase space. It provides an example where the diffusion can be directly inferred from the microscopic chaos.
\end{abstract}

\submitto{\jpa}
\maketitle

\section{Introduction}
The study of the long time asymptotics of interacting particles on a lattice sits at the foundation of classical statistical physics \cite{landau2013statistical}. In order to meet its predictions, it is always assumed that the underlying (microscopic) dynamics is chaotic, or at least ergodic.
Indeed ergodicity sits at the lowest level of the chaotic hierarchy in dynamical systems. 
It is customary to instead choose models with strong chaos to check the relevance of a statistical approach. For example the study of a classical uniformly hyperbolic map acting along a one-dimensional chain leads to an exact derivation of the diffusion coefficient \cite{gaspard_diffusion_1992}. More generally chaotic systems have been analysed using a statistical mechanics approach, see e.g.~\cite{Dorfman}. The required amount of chaos to apply such techniques remains a tantalising problem.
Even quantifying the amount of chaos for a given model can be subtle. One way is to look at how an initial perturbation at one site of the lattice spreads across the lattice with growing time. Ergodicity is achieved when the whole lattice is visited in the long time asymptotics, usually after averaging over initial conditions. Although intuitive, this definition can be less easy to check as the limits of large lattice size and long time do not commute in general.
Another way to investigate the presence of chaos, e.g. as performed in \cite{de2012largest} for classical spin chains, consists of computing the Lyapunov exponent of the trajectory of the chain when it is perturbed at one site at the initial time.
Another recent measure for classical chaos in many-body systems consists of studying the susceptibility fidelity \cite{PhysRevX.10.041017}.

The study of \emph{classical} spin chains has attracted a renewed interest in the last decade to understand how their quantum counterpart may reach or fail thermalisation. One ingredient called Out-of-Time-Order-Correlator (OTOC) introduced in \cite{OTOC_Larkin} has become a common tool to detect a quantum analogue of the butterfly effect. At the classical level the main (i.e. largest) Lyapunov exponent is proportional to the rate of exponential growth of the classical OTOC, see the review \cite{garcia-mata_out--time-order_2023} and the references therein. 
It is worth insisting that an exact estimate of the Lyapunov exponents for any trajectory is only possible numerically in an overwhelming number of cases.

Another connection comes surprisingly from black hole physics. When studying the black hole information paradox, see e.g. \cite{susskind2004introduction}, it has been conjectured that black holes are the fastest information scramblers \cite{sekino2008fast}. In order to understand those properties, it may be helpful to model a black hole with a classical maximally chaotic lattice.

In one dimension, a chain of fully chaotic maps has been already considered in \cite{gutkin2016classical}. The authors chose a particular coupling in order to be able to build a symbolic dynamics for the classical hyperbolic dynamics and list its periodic orbits. This work built upon the analysis for diffusive chains of one-dimensional maps \cite{pethel2006symbolic}. The model we are considering below is very similar to the family of models studied in \cite{gutkin2016classical} as it consists of a chain of interacting cat maps each acting on the two-dimensional torus.  One crucial difference is that the coupling  between cells is specific: it has a different range for the position and the momentum coordinates. It does not have either the local generic form obtained from \cite{pethel2006symbolic}. Another difference is that the coupling used in \cite{axenides2023arnol} appears more straightforward to generalise for a longer interaction range and for higher-dimensional lattice.

We are studying here the dynamical properties and the transport problem for a specific type of chain (or a lattice) of coupled Arnold cat maps. This chain was introduced recently \cite{axenides2023arnol} and will be denoted from now on as a Fibonacci chain of cat maps. It is maximally chaotic in the sense of being uniformly hyperbolic with a dense set of periodic orbits.
One crucial benefit of the Fibonacci chains of cat maps is that the whole Lyapunov spectrum is known analytically as already explained in \cite{axenides2023arnol}. The Cauchy problem is also exactly solvable.

The plan of the paper is as follows. The main definition is given in Sect.~\ref{defFchain}. We remind the reader of the main properties of the one-body cat map and introduce how to arrange them on a lattice. The dynamical properties of this lattice of cat maps are studied in Sect.~\ref{chaincat}. More particularly we address the location and counting of periodic orbits, the ergodic and hyperbolic features of the dynamics. Last a specific instance of the initial value problem is considered for a chain (one-dimensional lattice). 
Finally our results are discussed and some concluding remarks are drawn in the Sect.~\ref{discuss}.

\section{Definition of the model}
\label{defFchain}

\subsection{Single cat map}
Cat maps form an example of a discrete time dynamics displaying the strongest form of chaos. They were famously discussed in \cite{ArnoldAvez}, labelled as toral automorphisms. A point on the $2-$dimensional torus is located by a vector $\vec{x}$ with coordinates $(q,p)\in [0:1)^2$. The extremal values refer to the same point: \emph{e.g.} $q=0$ denotes the same point as $q=1$ for any fixed $p$. The time evolution is discrete and each step consists of applying a linear map on the vector $\vec{x}$:
\begin{equation}
   \vec{x}_{m+1} = M \vec{x}_{m}\ \  {\rm mod }\ 1, \label{catmapn1}
\end{equation}
where $\vec{x}_m$ denotes the location of a phase space point at time $m$. $M$ is a $2\times 2$ matrix with integer coefficients and unit determinant: $M\in$ SL$(2,\zZ)$.
Those conditions make the mapping well defined on the torus: 
$$M\left(\vec{x}+\vec{N}\right)= M\vec{x}\  {\rm mod }\ 1,\qquad \vec{N}\in\zZ^2,$$ 
and area preserving. For such a low dimensional system, the matrix is also symplectic:
\begin{equation*}
  M^T J M = J , \qquad
  J=\left(\begin{array}{cc}
    0 & 1 \\
    -1 & 0
  \end{array}\right) \ ,
\end{equation*}
such that the map (\ref{catmapn1}) becomes Hamiltonian.

As already mentioned above, the cat maps are known to show strong chaos. This refers to the fact that the cat maps are uniformly hyperbolic: at every point $\vec{x}$ of the phase space, the tangent space can be decomposed into an unstable manifold and a stable manifold. As they are acting on a compact phase space (the $2-$dimensional torus $\mathbb{T}^2$), they are also Anosov diffeomorphisms. The matrix $M$ possesses two eigenvalues $\lambda$, $1/\lambda$ where we chose $|\lambda| <1$. The corresponding eigenvectors are denoted by $\vec{v}_+$ and $\vec{v}_-$ respectively.
Then the stable manifold at a given point $\vec{x}$ on the torus is the line on the square representing the unit torus containing $\vec{x}$ and along the direction of $\vec{v}_+$, whereas the unstable manifold is the line containing $\vec{x}$ and along the direction of $\vec{v}_-$. On a compact phase space, uniform hyperbolicity implies mixing which implies ergodicity. 

The periodic orbits of the map are given by points with rational coordinates, \emph{i.e.} $\vec{x}\in \qQ^2$. Their period depends on the number theoretic properties of the entries of $\vec{x}$ and of the coeffficients of $M$ \cite{percival_arithmetical_1987,keating_asymptotic_1991}.

\subsection{Lattice of cat maps}

Following the steps described in \cite{axenides2023arnol} it is possible to define a dynamical system for a square lattice in $d$ dimension, with $L$ points along the linear side such that the uncoupled dynamics at each site is a cat map as in (\ref{catmapn1}). Note that the approach adopted in \cite{axenides2023arnol} consisted first of identifying the most general symplectic linear map acting on the $2L-$dimensional torus. Then one choice inspired by Fibonacci sequences was described. We will discuss only this choice in the case of the chain ($d=1$) to illustrate our main findings.

A point $\vec{X}_m$ at time $m$ in the $2L-$dimensional phase space now contains the canonical coordinates $q_m^{(l)},p_m^{(l)}$ where $l$ labels each of the $L$ sites of the chain. They are stored as follows
\begin{equation}
\vec{X}_m=\left(
  \begin{array}{c}
    q^{(1)}_{m} \\
                q^{(2)}_{m} \\
                \vdots      \\
                q^{(L)}_{m} \\
                p^{(1)}_{m} \\
                p^{(2)}_{m} \\
                \vdots      \\
                p^{(L)}_{m}
            \end{array}\right) \ . \label{defX}
\end{equation}

In \cite{axenides2023arnol} the lattice dynamics was built from Fibonacci sequences $(f_m)$. They are defined for an integer $k\ge 1$ by
\begin{equation*}
  f_0=0,\ f_1=1,\quad f_{m+1}= f_{m-1}+k\, f_m  ,\qquad m\ge 1\ .
\end{equation*}
In particular the most usual sequence is obtained for $k=1$. Those sequences are related to a linear mapping with integer coefficients via
\begin{equation*}
  \left(\begin{array}{c} f_{m} \\ f_{m+1} \end{array}\right)=A_k
  \left(\begin{array}{c} f_{m-1} \\ f_{m} \end{array}\right),\qquad  A_k=
  \left(\begin{array}{cc}
    0 & 1 \\
    1 & k
  \end{array}\right)\ .
\end{equation*}
After projection on the torus, this linear mapping leads to a cat map for:
\begin{equation*}
    \vec{x}_{m+1} =M \vec{x}_{m}\ \  {\rm mod }\ 1, \quad M=A_k^2= \left(\begin{array}{cc}
    1 & k \\
    k & 1+k^2
  \end{array}\right),\quad \vec{x}_m\left(\begin{array}{c}
    q_m\\ p_m \end{array}\right)\ \ .
\end{equation*}
The dependance of $M$ on $k$ was omitted. Notice that $M$ is symmetric hence has only real eigenvalues. It is symplectic and defines a hyperbolic map for $k\neq 0$.

Consider now $L$ such sequences $(f_m^{(l)})$ for $1\le l \le L$ with the same $k$. They can be 'coupled' along a chain with periodic boundary conditions, say via a nearest neighbour interaction in the following way
\begin{eqnarray}
  f_{m+1}^{(l)}= f_{m-1}^{(l)} + k\, f_m^{(l)} + g\left( f_m^{(l-1)}+ f_m^{(l+1)}\right)\ ,\ 1\le l \le L, \ m \ge 1,\label{coupledFibo}
\end{eqnarray}
where the real number $g$ plays the role of the interaction strength. This choice of the coupling is key to obtain a cat map on the $2L-$dimensional torus, \emph{i.e.} for the whole chain. Indeed the linear mapping corresponding to (\ref{coupledFibo}) is now
\begin{equation}
\left(
  \begin{array}{c}
    f^{(1)}_{m} \\
    f^{(2)}_{m} \\
    \vdots      \\
    f^{(L)}_{m} \\
    f^{(1)}_{m+1} \\
    f^{(2)}_{m+1} \\
    \vdots      \\
    f^{(L)}_{m+1}
  \end{array}\right)= A_{k,g}\left(
  \begin{array}{c}
    f^{(1)}_{m-1} \\
    f^{(2)}_{m-1} \\
    \vdots      \\
    f^{(L)}_{m-1} \\
    f^{(1)}_{m} \\
    f^{(2)}_{m} \\
    \vdots      \\
    f^{(L)}_{m}
  \end{array}\right),
\end{equation}
where $A_{k,g}$ is the $2L\times 2L$ block matrix
\begin{equation*}
  A_{k,g}=\left(\begin{array}{cc} 0 & \one_L \\ \one_L & C \end{array}\right) \ .
\end{equation*}
Here $\one_L$ stands for the $L\times L$ identity matrix while the last $L\times L$ block is defined as
\begin{equation}
  \label{defC}C=\left(\begin{array}{cccccc}
  {k}&{g}&0&...&0&{g}\\
  {g}&{k}&{g}&...&0&0\\
  0&{g}&{k}&...&0&0\\
  \vdots&\vdots&\vdots&\ddots&\vdots&\vdots\\
  {g}&0&0&...&{g}&{k}\end{array}\right)
\end{equation}
It is worth noticing that $C$ is a circulant matrix. The cat map for this 'Fibonacci chain' is defined by
\begin{equation}
  \vec{X}_{m+1} =   M \vec{X}_{m}\ \  {\rm mod }\ 1,\qquad M=A_{k,g}^2=
  \left(\begin{array}{cc} \one_L & C \\ C & \one_L + C^2 \end{array}\right) \ .\label{defFibchaincat}
\end{equation}
In order to define a map well defined on the torus, $M$ must have integer entries, which forces $g$ to be an integer.
One can also check that the obtained matrix is symmetric and symplectic:
\begin{equation*}
    M^T J M = J , \qquad
  J=\left(\begin{array}{cc}
    0 & \one_L \\
    -\one_L & 0
  \end{array}\right) \ .
\end{equation*}
The equations of motion can be rewritten as
\begin{eqnarray}
\hspace{-3cm}q_{m+1}^{(l)}=&q_m^{(l)}+ k p_m^{(l)}+ g\left[ p_m^{(l-1)}+p_m^{(l+1)}\right]
\label{eq_qm_full}\\
\hspace{-3cm}p_{m+1}^{(l)}=&  k q_m^{(l)}+ g\left[ q_m^{(l-1)}+q_m^{(l+1)}\right] +(k^2+1+2g^2) p_m^{(l)}+  2kg\left[ p_m^{(l-1)}+p_m^{(l+1)}\right]+g^2\left[ p_m^{(l-2)}+p_m^{(l+2)}\right]&
\label{eq_pm_full},
\end{eqnarray}
where the operation 'mod $1$' is implicit and $l$ is the site index ranging from $1$ to $L$ in a cyclic way: $l+L$ and $l$ refer to the same site.


\section{Fibonacci chain of cat maps}
\label{chaincat}

In this Section we will focus on three more specific questions related to the Fibonacci chain of cat maps (\ref{defFibchaincat}). First we will describe how the Hopf method to prove the ergodicity of a single cat map can be generalised to the chain case. Then we will remind the reader how to locate and count its periodic orbits as in \cite{axenides2023arnol}. Last, and more related to a physical situation, the solution of the initial value problem is given when the chain is perturbed at one site at the initial time.

\subsection{Ergodicity of the chain of cat maps: the Hopf method}

We are repeating here the steps described in chap.~8 of \cite{Dorfman} to show the ergodicity of a map. This method is referred to as the Hopf method.
We want to prove the following statement:\\
For every function $f$ continuous on the $2L-$dimensional torus $\mathbb{T}^{2L}$, one has the equality
\begin{equation}
  \label{ergodicity}
  \lim_{T\to\infty} \frac{1}{T}\sum_{m=0}^{T-1} f\left(M^m \vec{X}\right) 
  =\int_{\mathbb{T}^{2L}} f(\vec{X})\ud\mu \quad \textrm{ a. s.},\ \ud \mu=  \prod_{l=1}^L \ud q^{(l)} \ud p^{(l)}\ .
\end{equation}
This equality means physically that the average of any observable quantity along a trajectory coincides with the average of the same observable over the whole phase space, irrespectively of the initial point. This is sometimes dubbed as 'the time average equals the ensemble (or phase space) average'. The latter refers to the right hand side of (\ref{ergodicity}). The observable $f$ is averaged over the measure $\ud\mu$ invariant under the dynamics (\ref{defFibchaincat}). The acronym 'a.s.' stands for almost surely: in other words the equality holds for all points $\vec{X}$ except for a set of measure zero in the sense of $\mu$.

The proof is divided into two steps: the time average is shown to be the same if one reverses the arrow of time. Then it is used to show that the limit must be a constant, independent of the initial point of the trajectory and its integral coincide with the phase space average.

An essential ingredient is the Birkhoff theorem, see \emph{i.e.} \cite{ArnoldAvez} p.~16:\\
For every $f$ integrable on $\mathbb{T}^{2L}$, the function $f^+$ defined as
\begin{equation*}
  f^+(\vec{X})=  \lim_{T\to\infty} \frac{1}{T}\sum_{m=0}^{T-1} f\left(M^m \vec{X}\right)
\end{equation*}
exists for almost every $\vec{X}$, in the sense of the measure $\mu$ defined in (\ref{ergodicity}). Further this function is invariant under the mapping (\ref{defFibchaincat})
$$ f^+(M\vec{X})=f^+(\vec{X}) \ $$
and is integrable such that
$$ \int_{\mathbb{T}^{2L}} f^+(\vec{X})\ud \mu= \int_{\mathbb{T}^{2L}} f(\vec{X})\ud \mu \ . $$
Let us define the time reversed version of $f^+$ as
\begin{equation*}
  f^-(\vec{X})=  \lim_{T\to\infty} \frac{1}{T}\sum_{m=0}^{T-1} f\left(M^{-m} \vec{X}\right)
\end{equation*}
Considering the time reversed version of (\ref{defFibchaincat}), where $M$ is changed for $M^{-1}$ which also has integer coefficients, the Birkhoff theorem states that $f^-$ exists almost everywhere, is invariant under the map $M^{-1}$ and integrable. Its integral also coincides with the integral of $f$ over $\mathbb{T}^{2L}$. Note that $M$ is an isomorphism of the torus so being invariant under $M^{-1}$ is equivalent with being invariant under $M$. Therefore both $f^+$ and $f^-$ are invariant under $M$. Next define
\begin{equation*}
  \mathcal{A}_\epsilon=\left\{ \vec{X}\in \mathbb{T}^{2L} ,  f^+(\vec{X})-f^-(\vec{X}) > \epsilon \right\} \ . 
\end{equation*}
This set is invariant under the mapping (\ref{defFibchaincat}). We can use the Birkoff theorem for the mapping restricted to this set to get
\begin{equation*}
  \int_{\mathcal{A}_\epsilon} f^+(\vec{X})\ud\mu =  \int_{\mathcal{A}_\epsilon} f^-(\vec{X})\ud\mu= \int_{\mathcal{A}_\epsilon} f(\vec{X})\ud\mu\ .
\end{equation*}
Using the definition of $\mathcal{A}_\epsilon$, the triangle inequality
\begin{equation*}
0=\int_{\mathcal{A}_\epsilon} \left[f^+(\vec{X}) - f^-(\vec{X})\right] \ud\mu \ge \epsilon \mu(\mathcal{A}_\epsilon)  
\end{equation*}
forces $\mu(\mathcal{A}_\epsilon)$ to be $0$. The same line of arguments can be used to prove that the set over which $f^--f^+$ is larger than a fixed $\epsilon$ is of zero measure. Hence $f^+$ and $f^-$ are equal almost everywhere. This ends the first part of the proof of the ergodicity of (\ref{defFibchaincat}).

Next we shall prove the following
\begin{itemize}
\item $f^+(\vec{X})$ does not depend on the component of $\vec{X}$ along any eigenvectors $\vec{v}_+^i$ associated to an eigenvalue $\lambda_i$ of $M$ obeying $|\lambda_i|<1$,
\item $f^-(\vec{X})$ does not depend on the component of $\vec{X}$ along any eigenvectors $\vec{v}_-^i$ associated to an eigenvalue $1/\lambda_i$ of $M$ obeying $|1/\lambda_i|>1$.
\end{itemize}
Together with the first part, this will end the proof that the $f^+$ is constant almost everywhere, hence equals its phase average\footnote{For our parametrisation of the torus, we have $\mu(\mathbb{T}^{2L})=1$.}.

To prove the statement about $f^+(\vec{X})$, one may proceed as follows. Consider a point $\vec{X}$ in the phase space and the neighbouring position at
\begin{equation*}
  \vec{W}=\vec{X}+\delta\ \vec{v}_+^{\, i}, \quad \delta=-\frac{\vec{X}.\vec{v}_+^{\, i}}{|\vec{v}_+^{\, i}|}\ .
\end{equation*}
By definition of $\vec{v}_+^{\, i}$
$$ M \vec{v}_+^{\, i}=\lambda_i \vec{v}_+^{\, i}$$
Hence
$$ | M^m\vec{W}-M^m\vec{X}| = | \delta\ M^m\vec{v}_+^{\, i} | = |\delta|\, |\lambda_i|^m\, |\vec{v}_+^{\, i}| \to 0, \ m\to +\infty\ . $$
Using the continuity of $f$
\begin{equation}
  \lim_{m\to\infty} M^m\vec{X}= \lim_{m\to\infty} M^m\vec{W} \Rightarrow \lim_{m\to\infty} f(M^m\vec{X})= \lim_{m\to\infty} f(M^m\vec{W})\ .\label{fwfx}
\end{equation}
For the definition of $f^+$, the sum can be divided as follows
\begin{equation*}
  \frac{1}{T}\sum_{m=0}^{T-1} f\left(M^m \vec{W}\right)= \frac{1}{T} \left[ \sum_{m=0}^{T_\epsilon-1} f\left(M^m \vec{W}\right)+ \sum_{m=T_\epsilon}^{T-1} f\left(M^m \vec{W}\right) \right]
\end{equation*}
From (\ref{fwfx}) the second term can be made arbitrarily close to the same sum evaluated at $\vec{X}$ instead of $\vec{W}$ for a given $T_\epsilon$. For a large enough $T$ the first term can be also be made small enough so that
$$ f^+(\vec{W})=\lim_{T\to\infty} \frac{1}{T}\sum_{m=0}^{T-1} f\left(M^m \vec{W}\right) =f^+(\vec{X})\ . $$
From the definition of $\vec{W}$ this shows that $f^+(\vec{X})$ does not depend on the component of $\vec{X}$ along the direction $\vec{v}_+^{\, i}$. This is true for any choice of $i$ hence $f^+(\vec{X})$ remains constant if $\vec{X}$ is translated anywhere within the stable manifold. Similarly if one defines ($i$ is fixed)
\begin{equation*}
  \vec{Z}=\vec{X}+\delta\ \vec{v}_-^{\, i}, \quad \delta=-\frac{\vec{X}.\vec{v}_-^{\, i}}{|\vec{v}_-^{\, i}|}\ .
\end{equation*}
it can be proved that $f^-(\vec{Z})=f^-(\vec{X})$ hence $f^-$ does not depend on the component of $\vec{X}$ along the direction $\vec{v}_-^{\, i}$. This is valid for every $i$ so $f^-(\vec{X})$ does not depend on the components of $\vec{X}$ along its unstable manifold.

Using that $f^+$ and $f^-$ are equal almost everywhere and that the eigenvectors of $M$ form a complete set, we can conclude that $f^+=f^-$ is a constant function almost everywhere on the torus. Its value can be estimated from its integral over the torus, which finishes the proof of (\ref{ergodicity}).

\subsection{Numerical validation of ergodicity}

The results of the previous section can be illustrated via a numerical experiment showing that the ensemble average and the time average coincide for system (\ref{defFibchaincat}). 
In anticipation for the transport problem below, the set of possible initial conditions is restricted to be:
\begin{equation}
  q_0^{(l)}= q_0 \delta_{l,(L{+}1)/2},\qquad p_0^{(l)}=0\ , \label{IC_chain}
\end{equation}
where $\delta_{l,m}$ stands for the Kronecker delta and $q_0\in [0:1)$. The ensemble average consists of an average over random uniformly distributed $q_0$.

Spacetime diagrams below displays the propagation of an initial propagation of a perturbation at the centre of the chain. The horizontal axis shows the time while the vertical axis stands for the position along the chain. As shown below, the perturbation reaches the chain's ends in a time proportional to the chain's size. Therefore our numerics were always performed for the duration
$$ 1 \le m \le \frac{L+7}{4}\ .$$
In particular taking $m$ close or equal to the upper bound defines the long time limit of our chain. It diverges in the limit of the large chain's size $L\to\infty$.\\
Each of the spacetime diagrams contains $({L(L+7)})/{4}$ cells. The color code for each cell is the value of the position vector at that time $q_m^{(l)}$.

\begin{figure}[H]
\centering
\begin{subfigure}{0.9\subfigwidth}
  \centering
  \includegraphics[width=1\linewidth]{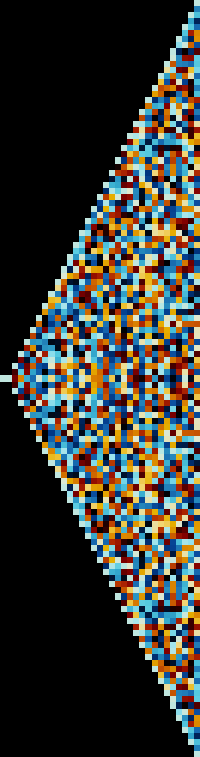}
  \caption{}
\end{subfigure}
\begin{subfigure}{0.9\subfigwidth}
  \centering
  \includegraphics[width=1\linewidth]{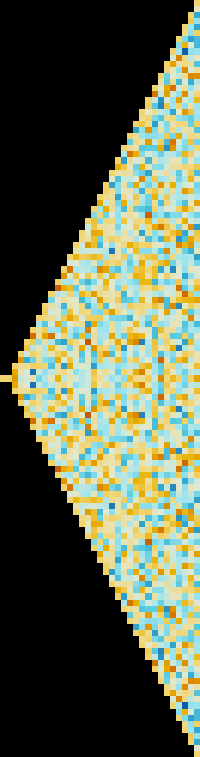}
  \caption{}
\end{subfigure}
\begin{subfigure}{0.9\subfigwidth}
  \centering
  \includegraphics[width=1\linewidth]{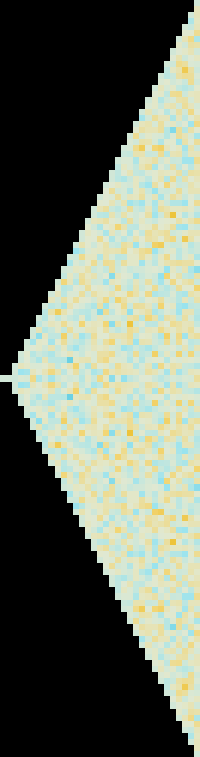}
  \caption{}
\end{subfigure}
\begin{subfigure}{0.9\subfigwidth}
  \centering
  \includegraphics[width=1\linewidth]{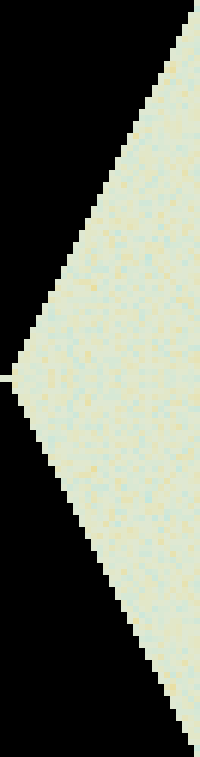}
  \caption{}
\end{subfigure}
\begin{subfigure}{0.9\subfigwidth}
  \centering
  \includegraphics[width=1\linewidth]{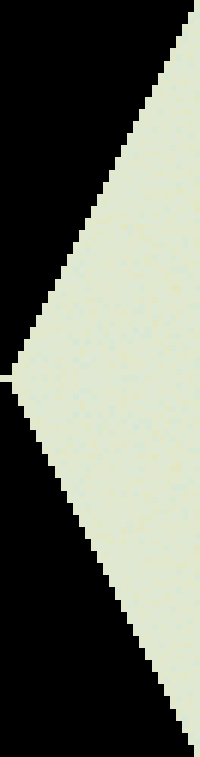}
  \caption{}
\end{subfigure}
\begin{subfigure}{0.38\subfigwidth}
  \centering
  \includegraphics[width=1\linewidth]{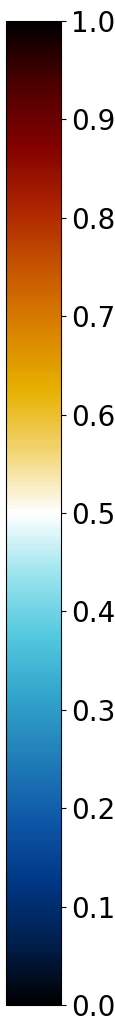}
  \caption{}
\end{subfigure}
\caption{Space-time diagrams for the mapping (\ref{defFibchaincat}) with the initial condition (\ref{IC_chain}) for $k=1$, $g=1$, $L=125$. The horizontal axis stands for the time $m$ whereas the vertical axis stands for the position along the chain $l$. Various levels of sampling are shown, illustrating convergence towards the equilibrium state when the number of sampled initial conditions grows: $1, 10, 10^2, 10^3, 10^4$ going from left (a) to right (e) respectively. (f) shows the the meaning of the colours.}
\label{Ensemble Average}
\end{figure}

First, an ensemble average was performed for the initial condition (\ref{IC_chain}). This amounts to sampling randomly the values of $q_0$ uniformly in $[0,1)$. Fig.~\ref{Ensemble Average} illustrates that the ensemble average converges for a large enough amount of initial conditions to a constant  profile in the long time limit as defined above, see more details about the used computational method in \ref{comput_methods}.

A time average can be also performed numerically to mimic the left hand side of (\ref{ergodicity}). We perform a sliding average following
\begin{equation}
  f_{\Delta m}(\vec{X},m)=\frac{1}{2\Delta m}\sum_{j=m-\Delta m}^{m+\Delta m} f\left( M^j \vec{X}\right) \label{slid_t_av}
\end{equation}
and consider it for fixed $\Delta m$ such that $1\ll \Delta m\ll m$ for large $m$. 


\begin{figure}[H]
\centering
\begin{subfigure}{\subfigwidth}
  \centering
  \includegraphics[width=1\linewidth]{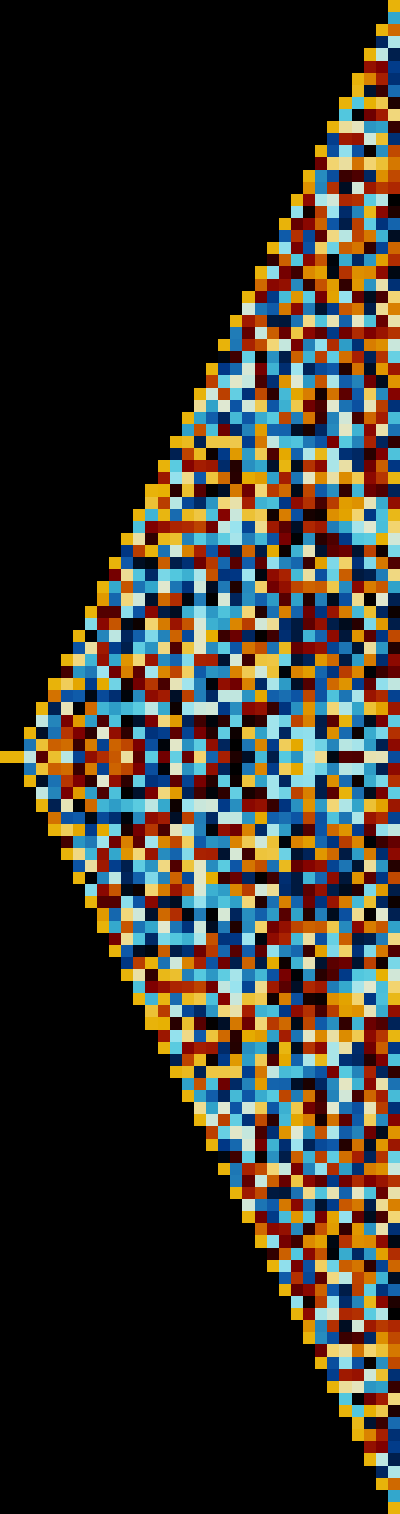}
  \caption{}
\end{subfigure}
\begin{subfigure}{\subfigwidth}
  \centering
  \includegraphics[width=1\linewidth]{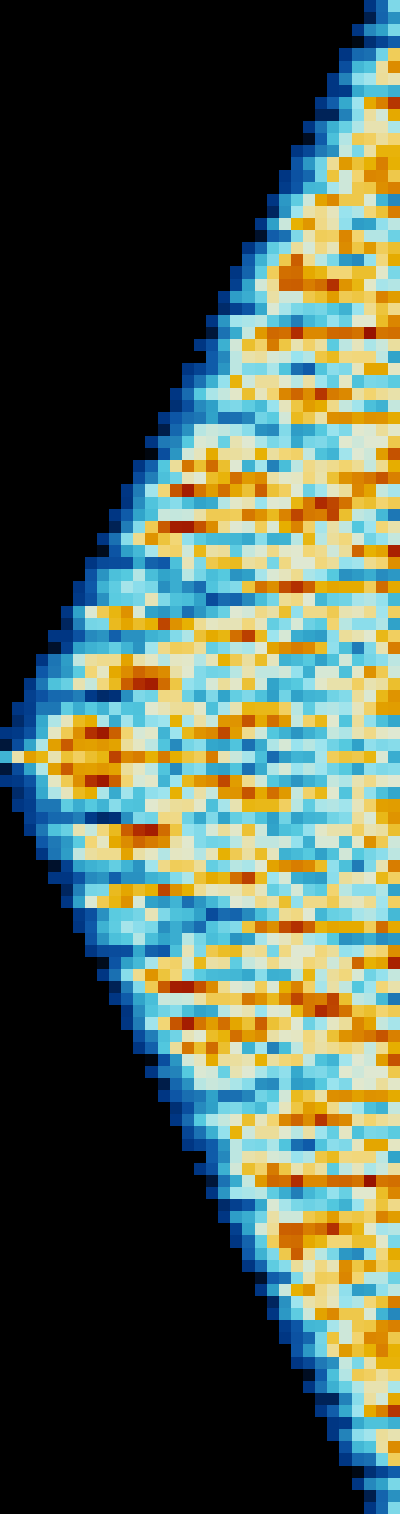}
  \caption{}
\end{subfigure}
\begin{subfigure}{\subfigwidth}
  \centering
  \includegraphics[width=1\linewidth]{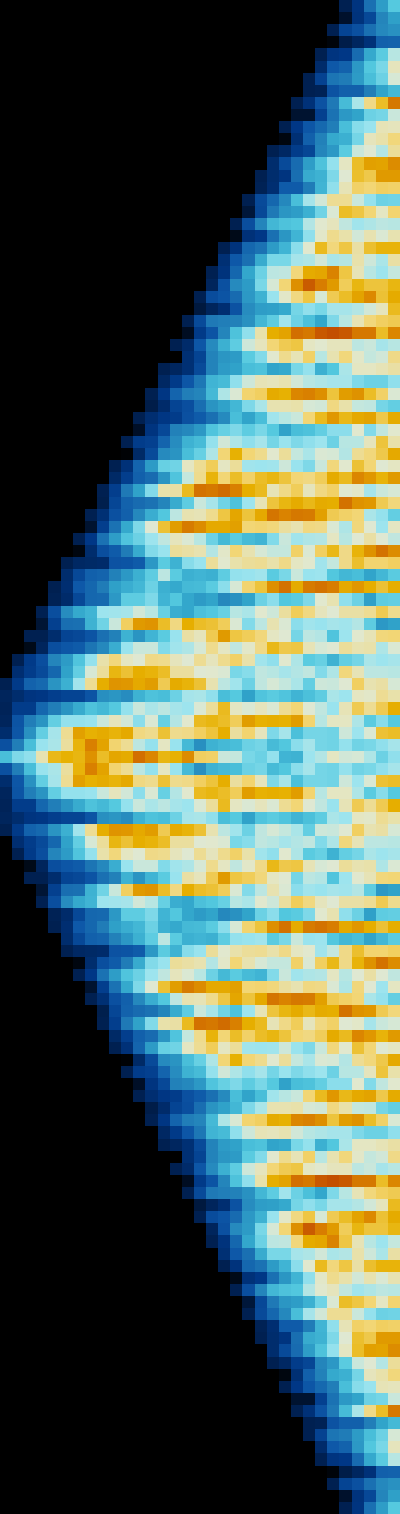}
  \caption{}
\end{subfigure}
\begin{subfigure}{\subfigwidth}
  \centering
  \includegraphics[width=1\linewidth]{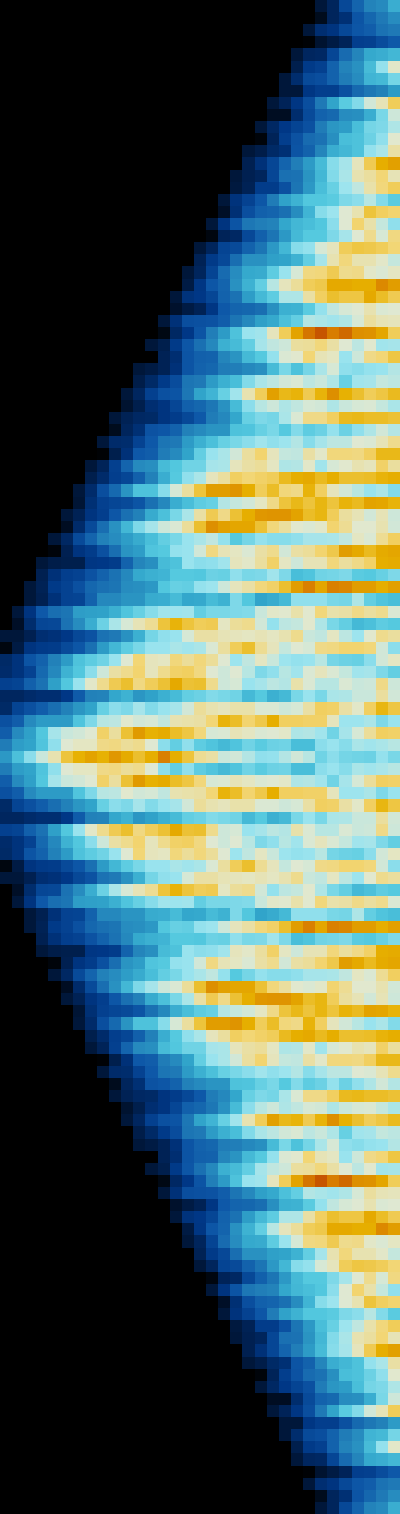}
  \caption{}
\end{subfigure}
\begin{subfigure}{\subfigwidth}
  \centering
  \includegraphics[width=1\linewidth]{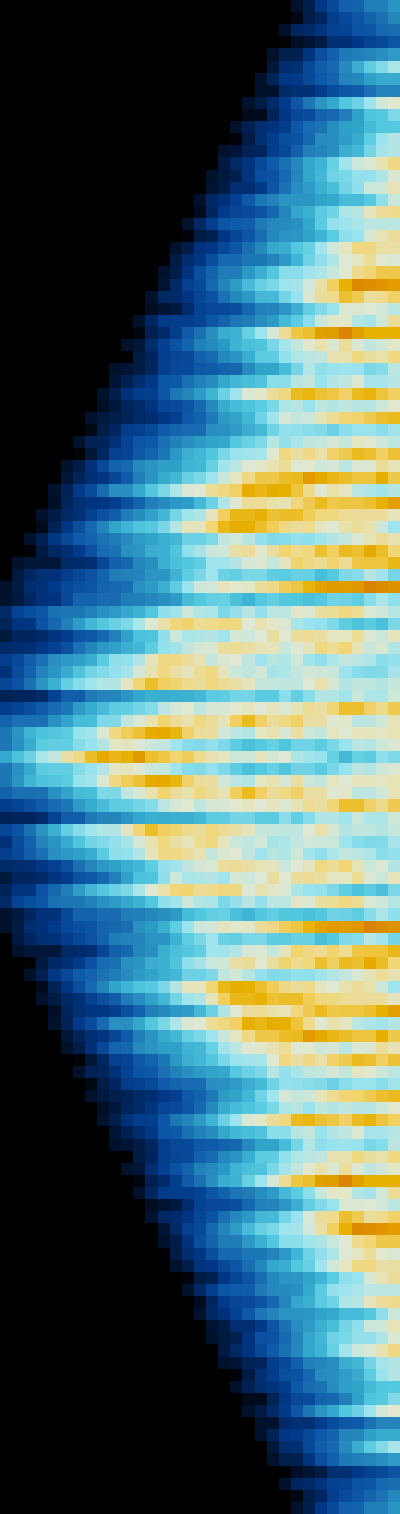}
  \caption{}
\end{subfigure}
\caption{Space-time diagrams for the mapping (\ref{defFibchaincat}) with the initial condition (\ref{IC_chain}) for $k=1$, $g=1$, $L=125$, $q_0=\frac{\sqrt{5}-1}{2}$ across various levels of time averaging, illustrating convergence towards the equilibrium state for $\Delta m=1, 5, 9, 13, 17$ going from left (a) to right (e) respectively.}
\label{Time Average}
\end{figure}
\begin{figure}[!ht]
     \begin{minipage}[left]{0.49\linewidth}
     \begin{center}
       \includegraphics[width=0.5\linewidth]{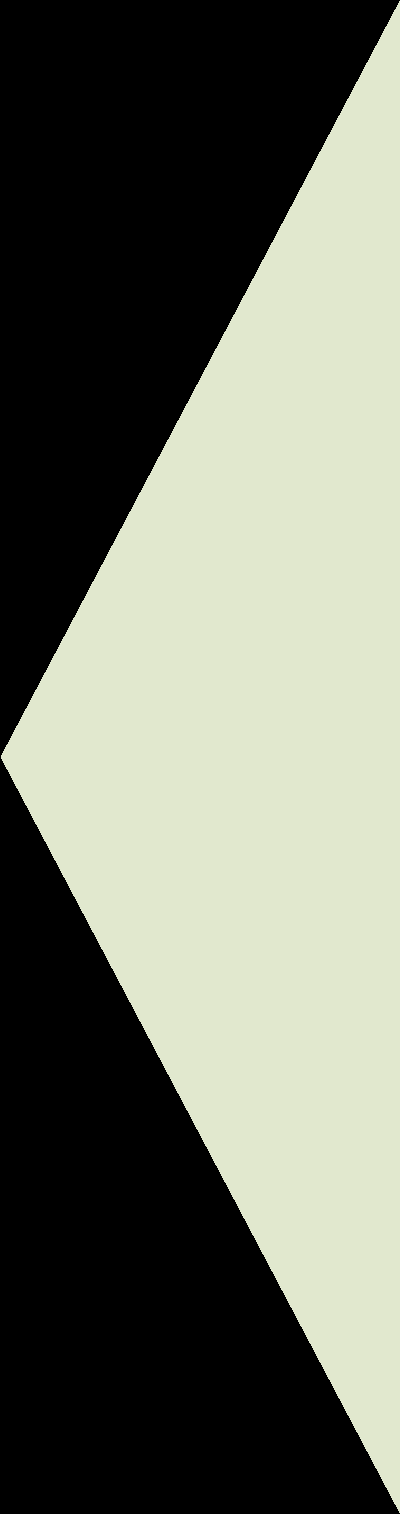}
     \end{center}
    \end{minipage}
    \begin{minipage}[right]{0.49\linewidth}
      \begin{center}
       \includegraphics[width=0.5\linewidth]{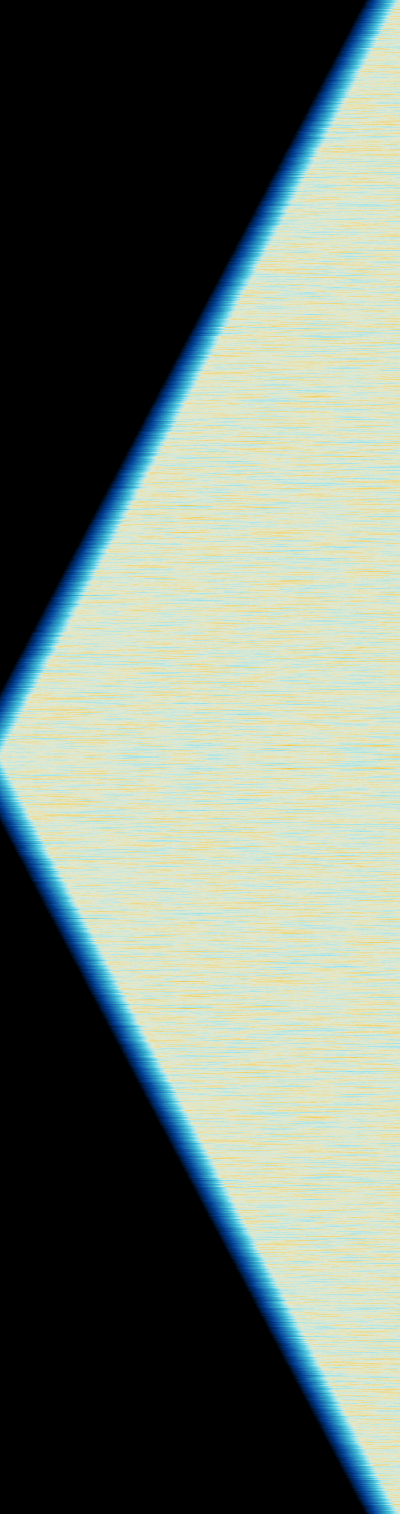}
     \end{center}
    \end{minipage}
    \caption{Comparison of the ensemble average (left) and time average (right) for a chain $L=10 001$, $k=1$, $g=1$. On the left it is averaged over $10^4$ initial randomly distributed values. On the right, $q_0=\frac{\sqrt{5}-1}{2}$ and the width of the averaging window was taken to be $\Delta m=201$.}\label{ergo_large_size}
\end{figure}

Figure~\ref{Time Average} illustrates that this time average leads to the same constant profile as in Figure~\ref{Ensemble Average}.
It was checked that the structure of the time average is independent of the choice of the initial perturbation $q_0$. The agreement between ensemble and time average becomes clearer for larger sizes as illustrated in Fig.~\ref{ergo_large_size}.

\subsection{Counting the periodic orbits of the map}

From now on, we shall assume that {$L$ is a positive odd integer}. This means that the matrix $M$ in (\ref{defFibchaincat}) does not have a unit eigenvalue \cite{axenides2023arnol}. Given that the dynamical system (\ref{defFibchaincat}) is linear, it is easier to locate and count its periodic orbits.

A phase space point belongs to a periodic orbit of period $m>0$ if and only if
\begin{equation*}
  M^m\vec{X}=\vec{X} {\rm mod}\ 1 \textrm{ and } M^{m-1}\vec{X}\neq\vec{X} {\rm mod}\ 1\ {\rm when }\ m\ge 1 .
\end{equation*}
For any interaction strength $g\ge 1$, the only periodic orbit of period one, aka a fixed point, of $M$ is the point $\vec{0}$. A point along a periodic orbit of period $m$ is a solution of
\begin{equation*}
  M^m\vec{X}=\vec{X}+\vec{N}, \quad \vec{N}\in\zZ^{2L}\ .
\end{equation*}
As $1$ is not in the spectrum of $M$ and this spectrum is real, the matrix $M^m-\one_{2L}$ is invertible for any $m\ge 1$. Hence a point along a periodic orbit is always given by
\begin{equation*}
  \vec{X}=B^{-1}\vec{N}\ ,\  B=\left( M^m -\one_{2L}\right)\ .
\end{equation*}
Using the fact that $B$ has integer entries and its inverse can be written as a sum over minors, we can see immediately that the periodic orbits of the cat map (\ref{defFibchaincat}) are the points with rational coordinates. They form a dense set of zero measure in the phase space $\mathbb{T}^{2L}$.

The search for the periodic orbits of exact period $m$ amounts for solving 
$$ B\vec{X}=\vec{0}\ {\rm mod}\ 1\ . $$ 
The number of solutions of this congruence equation $N(m)$ can be computed using the elementary divisors of $B$, see Appendix A.23 in \cite{Mehta}. As $B$ is a matrix with integer entries, some elementary operations (interchanging rows and columns, changing the sign of a row or a column, adding an integer multiple of a row- to any other row) can be performed to rewrite $B$ as
\begin{equation}
  B=P^{-1} D Q^{-1}, \label{elemdivB}
\end{equation}
where $P$ and $Q$ are matrices with integer entries and determinant equal to $\pm 1$. $D$ is a diagonal matrix
\begin{equation*}
  D=\left(\begin{array}{ccccc}
  {d_1}&{0}&0&...&0\\
  {0}&{d_2}&{0}&...&0\\
  0&{0}&{d_3}&...&0\\
  \vdots&\vdots&\vdots&\ddots&\vdots\\
  {0}&0&0&...& d_{2L}\end{array}\right)\ ,
\end{equation*}
such that $d_i$ is a divisor of $d_{i+1}$ for $1\le i \le 2L-1$. This decomposition is also sometimes called Smith normal form. The form (\ref{elemdivB}) enables one to count the solutions of the equation $ B\vec{X}=\vec{0}\ {\rm mod}\ 1$ in a simple way:
\begin{eqnarray*}
  B\vec{X}\equiv 0 \ {\rm mod}\ 1 &\iff& \exists \vec{N}\in \zZ^{2L},\ P^{-1} D Q^{-1}\vec{X}=\vec{N}\ ,\\
  &\iff& \exists \vec{N}\in \zZ^{2L},\ DQ^{-1}\vec{X}=P\vec{N}\ .\\
\end{eqnarray*}
Now use the fact that $P$,$Q$ are invertible so that the solutions of the last equation are in one-to-one correspondence with the solutions $\vec{y}$ of
$$ D\vec{y}=\vec{N}', \quad \vec{N}' \in \zZ^{2L} \iff D\vec{y}\equiv 0 \ {\rm mod}\ 1\ . $$
The system is now decoupled and one can count the solutions $y_i$ for each equation $d_i y_i\equiv 0$ mod $1$ separately. They consist of the rational numbers $k/d_i$ for $0\le k\le |d_i|-1$ hence there are exactly $|d_i|$ solutions. The total number of solutions of $D\vec{y}\equiv 0$ is then
$$ \prod_{i=1}^{2L} |d_i|= |\det D| = |\det B|\ ,$$
where (\ref{elemdivB}) was used in the last equality.\\
Calling $N(m)$ the number of all periodic orbits of period $m$, including repetitions, we have found
\begin{equation}
  \label{exactNm}
  N(m)=\left| \det(M^m-\one_{2L})\right| \ ,
\end{equation}
which generalises the formula known for one cat map, see e.g. \cite{Dorfman}. This exact formula is helpful to determine the topological entropy $h_{top}$ of the Fibonacci lattice of cat maps. It can be defined by the asymptotic formula
\begin{equation}
  \label{defhtop}
  N(m)\sim m^{a} e^{h_{top} m}, m\to \infty \ ,
\end{equation}
where $a$ is an arbitrary real number.\\
For our particular lattice model, all the eigenvalues of $M$ are known exactly, hence one has an exact formula for $N(m)$. They were already explicited in \cite{axenides2023arnol} so we shall only remind the reader of the main steps, see also \ref{IVP}. The matrix $C$ defining $M$ in (\ref{defFibchaincat}) is a circulant matrix hence can be diagonalised by the discrete Fourier transform. Its eigenvalues are in the case of a chain with only nearest neighbour interaction:
\begin{equation}
  \label{eigvC}
  D_l=k+2g\cos\left(\frac{2\pi l}{L}\right), \quad 0\le l\le L-1\ .
\end{equation}
Due to the block structure of $M$, each eigenvalue $D_l$ of $C$ leads to a couple of eigenvalues $(\lambda_l,1/\lambda_l)$ of $M$. Choosing, say, $0< \lambda_l<1$, those eigenvalues are the roots of
$$ X^2 -(2+D_l^2) X +1 = 0 $$
from which one easily finds
$$\lambda_l=\frac{2+D_l^2}{2}-|D_l|\frac{\sqrt{D_l^2+4}}{2} . $$
This leads to the exact expression for the Lyapunov exponents of (\ref{defFibchaincat}): $\pm\Lambda_l$ with $\Lambda_l=|\ln\lambda_l|$, as already discussed in \cite{axenides2023arnol}. From this explicit expression of the eigenvalues of $M$, one can deduce
\begin{eqnarray*}
  N(m)&= \left| \prod_{l=1}^L \left( \lambda_l^m-1\right)\left( \frac{1}{\lambda_l^m}-1\right)\right|=&\prod_{l=1}^L \left[2\cosh(m \Lambda_l) -2\right], \quad \Lambda_l=|\ln\lambda_l|\ .
\end{eqnarray*}
The asymptotic for large $m$ is
\begin{equation*}
  N(m)\sim \prod_{l=1}^L \exp\left(m \Lambda_l\right)\sim \exp\left( m \sum_{l=1}^L \Lambda_l \right) \ ,\quad m\to \infty\ ,
\end{equation*}
which means that the topological entropy coincides with the sum of the positive Lyapunov exponents. From \cite{axenides2023arnol}, this entropy coincides with the invariant Kolmogorov-Sinai entropy:
\begin{equation*}
h_{top}=h_{KS}=\sum_{l=1}^L \Lambda_l,\qquad \Lambda_l=\left|\ln\left(\frac{2+D_l^2}{2}-|D_l|\frac{\sqrt{D_l^2+4}}{2}\right)\right|\ .
\end{equation*}
This is another instance of Pesin's identity, see e.g.~\cite{RevModPhys.57.617}. 

\section{Transport problem}

\subsection{Definition}
It was already highlighted in \cite{axenides2023arnol} that the Initial Value Problem (aka Cauchy's problem) for (\ref{defFibchaincat}) can be solved exactly, see a reminder in \ref{IVP}. We shall use those results to focus on the propagation on an initial perturbation of the chain. 
The formul\ae{} (\ref{IVP_q}) and (\ref{IVP_p}) will become more explicit for the following initial conditions:

\begin{equation}
    q_0^{(l)} = q_0\; \delta_{l,\frac{L+1}{2}},\qquad p_0^{(l)}=0 \ ,\label{init_localpert}
\end{equation}
For simplicity, we shall assume from now on that
$$ m\ge 1\ .$$

\subsection{Growth of the profile support}
Using from the definition (\ref{defmatF}) that
$$ F^\dag\vec{p}_0=\vec{0},\quad \left(F^\dag \vec{q}_0\right)^{(l)}= q_0 \frac{e^{-2\pi\ic l (L+1)/2L}}{\sqrt{L}}$$
the position at later times of the chain is given by a single sum
\begin{equation}
    {q}_m^{(l)}= q_0 \frac{1}{L} \sum_{s=1}^L A^{(m)}_s e^{\frac{2\pi\ic s}{L} \left[l-  \frac{L+1}{2}\right] } ,\label{qm_localpert}
\end{equation}
where the coefficients $A^{(m)}_s$ are defined by
\begin{equation}
    A^{(m)}_s= -\frac{\lambda_{s}^{m-1}- \lambda_{s}^{-m+1}}{\lambda_{s}- \lambda_{s}^{-1}}+ \frac{\lambda_{s}^{m}- \lambda_{s}^{-m}}{\lambda_{s}- \lambda_{s}^{-1}}\label{defAm}
\end{equation}

Our main observation, using a similar way of thinking as in Appendix A of \cite{axenides2023arnol}, consists of noticing that the expression appearing in (\ref{defAm}) is a polynomial with integer coefficients of $D_s^2$ defined by
\begin{equation}
     D_s=k+2g \cos\left(\frac{2\pi s}{L}\right) \ .\label{defDl}
\end{equation}
Specifically
\begin{equation}
    \frac{\lambda_{s}^{m}- \lambda_{s}^{-m}}{\lambda_{s}- \lambda_{s}^{-1}}=P_m(D_s^2), \label{defPm}
\end{equation}

where $P_m$ is the polynomial
\begin{equation}
    P_1(X)=1, \quad P_m(X)= X^{m-1} + \displaystyle\sum_{j=0}^{m-2} {m+j \choose 2j+1} X^j,m\ge 2\ . \label{coeffPm}
\end{equation}
A way to see it is to revisit (\ref{eq_rm}) and rewrite it as a recursion relation for a family of polynomials. Let one define such polynomials as 
\begin{equation}
    P_0(X)=0, \ P_1(X)=1,\quad P_{m+2}(X)=(X+2)P_{m+1}(X) -P_m(X),\ m\ge 0\ .\label{recurrPm}
\end{equation}
These polynomials are monic with integer coefficients. Besides
\[ \textrm{deg }P_m= m-1, m\ge 1 \ .\]
If ones decides to solve (\ref{eq_rm}) with an ansatz as above $ r_{m}^{(l)}=a_l \lambda_l^m$ and search for the solution obeying $r_0^{(l)}=0$, $r_1^{(l)}=1$ one finds
$$  r_{m}^{(l)}=\frac{\lambda_{l}^{m}- \lambda_{l}^{-m}}{\lambda_{l}- \lambda_{l}^{-1}}\ .$$
By unicity of the solution this has to be also $P_m(D_l^2)$, which proves (\ref{defPm}). For instance, the first polynomials are
\begin{eqnarray*}
\hspace{-2cm}     P_1(X)=&& 1\\
\hspace{-2cm}     P_2(X)=&& X+2\\
\hspace{-2cm}     P_3(X)=&& X^2+4X+3\\
\hspace{-2cm}     P_4(X)=&& X^3+6X^2+10X+4\\
\hspace{-2cm}     P_5(X)=&& X^4+8X^3+21X^2+20X+5\\
\hspace{-2cm}     P_6(X)=&& X^5+10X^4+36X^3+56X^2+35X+6\\
\hspace{-2cm}     P_7(X)=&& X^6+12X^5+55X^4+120X^3+126X^2+56X+7\\
\hspace{-2cm}     P_8(X)=&& X^7+14X^6+78X^5+220X^4+330X^3+252X^2+84X+8\\
\hspace{-2cm}     P_9(X)=&& X^8+16X^7+105X^6+364X^5+715X^4+792X^3+462X^2+120X+9\\
 \hspace{-2cm} P_{10}(X)=&& X^9+18X^8+136X^7+560X^6+1365X^5+2002X^4+1716X^3+792X^2+165X+10
\end{eqnarray*}
Finally a closer look at the coefficients leads to the expansion (\ref{coeffPm}).
\bigskip

We want to prove that the initial condition (\ref{init_localpert}) will lead to a profile of nonzero values along the chain, whose support has a length which is increasing with the time $m$. From the form of the position profile (\ref{qm_localpert}) the edges of the profile are given by the contribution of $A^{(m)}_s$ containing the highest (modulo $L$) Fourier harmonics. From (\ref{defAm}), (\ref{defPm}) and (\ref{coeffPm}) one can expand $A^{(m)}_s$ for large $D_s$
\[ A^{(m)}_s=-P_{m-1}(D_s^2)+P_m(D_s^2)\approx D_s^{2(m-1)} +(2m-3) D_s^{2(m-2)} +\dots,\quad D_s\gg 1\ .\]
From the definition (\ref{defDl}), the expansion of $A^{(m)}_s$ into decreasing powers starts with
\begin{equation*}
 A^{(m)}_s \approx (2g)^{2m-2}\cos^{2m-2}\left(\frac{2\pi s }{L}\right)+(2m-2)k (2g)^{2m-3} \cos^{2m-3}\left(\frac{2\pi s }{L}\right)+\dots\ .
\end{equation*}
Using the linearisation formul\ae{}
\[ \hspace{-2cm}\cos^{2\alpha}\theta=\frac{1}{2^{2\alpha-1}} \left[ \frac{1}{2} {2\alpha \choose \alpha}+\sum_{r=1}^\alpha {2\alpha \choose \alpha-r}\cos(2 r \theta) \right], \quad \cos^{2\alpha+1}\theta=\frac{1}{2^{2\alpha}} \sum_{r=0}^\alpha {2\alpha+1 \choose \alpha-r}\cos[(2 r+1) \theta]\ ,\]
the expansion of $A^{(m)}_s$ into decreasing harmonics is
\begin{equation*}
    A^{(m)}_s \approx \frac{(2g)^{2m-2}}{2^{2m-3}}\cos\left(\frac{2\pi s(2m-2) }{L}\right) + \frac{(2m-2)k (2g)^{2m-3}}{2^{2m-4}}\cos\left(\frac{2\pi s(2m-3) }{L}\right)   +\dots\ .
\end{equation*}
The formula for the positions along the chain becomes
\begin{eqnarray}
    {q}_m^{(l)}\approx  q_0 \Big\{ 2g^{2m-2}\frac{1}{L} \sum_{s=1}^L \cos\left(\frac{2\pi s(2m-2) }{L}\right) e^{\frac{2\pi\ic s}{L} \left[l-  \frac{L+1}{2}\right] } + \nonumber\\ 
    2k(2m-2) g^{2m-3} \frac{1}{L} \sum_{s=1}^L \cos\left(\frac{2\pi s(2m-3) }{L}\right) e^{\frac{2\pi\ic s}{L} \left[l-  \frac{L+1}{2}\right] } +\dots \Big\}\ .
\end{eqnarray}
Now using
\begin{equation}
 \frac{1}{L} \sum_{s=1}^L \cos\left(\frac{2\pi s r }{L}\right) e^{\frac{2\pi\ic s}{L} \left[l-  \frac{L+1}{2}\right] }=\frac{1}{2} \delta_{l,\frac{L+1}{2}-r}+\frac{1}{2} \delta_{l,\frac{L+1}{2}+r},   \label{orthocosexp}
\end{equation}
one gets finally
\begin{equation}
 \hspace{-3cm}{q}_m^{(l)}\approx  q_0 g^{2m-2} \left[\left( \delta_{l,\frac{L+1}{2}-(2m-2)}+\delta_{l,\frac{L+1}{2}+(2m-2)} \right) +  (2m-2)\frac{k}{g}\left( \delta_{l,\frac{L+1}{2}-(2m-3)}+\delta_{l,\frac{L+1}{2}+(2m-3)} \right)+\dots \right]  \ . \label{qml_edges}
\end{equation}
The position profile is therefore supported on a light cone whose ends are located at the sites 
\begin{equation}
    l_{\pm}= \frac{L+1}{2}\pm (2m-2)\ , m\ge 1,\label{def_light_cone}
\end{equation}
so that its width $l_+-l_-$ grows linearly in time:   $l_+-l_-=4(m-1)$. 
The value of the position at the edges of the profile is
\begin{equation}
    q_{m}^{(l_\pm)}= q_0 g^{2m-2}\ {\rm mod }\ 1 \ .\label{qedge_q0init}
\end{equation}
When the range of interaction is changed for $r\ge 1$ along the chain, the ends of the light cone for the position profile become:
\begin{equation}
    l_\pm=\frac{L+1}{2}\pm 2r(m-1)\ .\label{def_light_cone_arbitrary_r}
\end{equation}

\subsection{Approximating the chain profile in the long time}
\label{profile_asympt}

The exact analytic expression of the profile enables one to build an approximation at long time, to be compared with the numerics. While the last steps are heuristic, this approach may be helpful for a more general intuition and will be further justified by numerics.

The position profile at time $m$ and site $l$ was already seen to be given by 
\begin{equation}
{q}_m^{(l)}= \frac{q_0}{L} \sum_{s=1}^L\left[P_m(D_s^2)-P_{m-1}(D_s^2)\right] e^{\frac{2\pi\ic s}{L} \left[l-  \frac{L+1}{2}\right] }, \label{profile_qm}
\end{equation}
with $D_s=k+2g\cos(2\pi s/L)$. Let us now consider the function
\begin{equation}
F_m(s)=   P_m(D_s^2)-P_{m-1}(D_s^2)\ .\label{defFm} 
\end{equation}
As it is a real polynomial function of $\cos({2\pi s}/{L})$, $F_m(s)$ is a real even $L-$periodic function of $s$, which is now seen as a continuous variable. It can be expanded as a Fourier series 
\begin{equation}
    F_m(s)=\sum_{p=0}^\infty f^{(m)}_p \cos\left(\frac{2\pi p s}{L}\right)\ .\label{FserieFm}
\end{equation}
with the Fourier coefficients given by
\begin{equation}
    f_0^{(m)}= \frac{1}{L} \int_0^L F_m(s) \ud s,\quad \forall p\ge 1, \ 
    f_p^{(m)}= \frac{2}{L} \int_0^L F_m(s) \cos\left(\frac{2\pi p s}{L}\right)\ud s
\end{equation}
It may be worth noticing that $F_m(s)$ is from its definition (\ref{defFm}) a polynomial of finite powers of trigonometric functions so its Fourier series (\ref{FserieFm}) actually has always a finite number of nonzero coefficients for finite $L,m$.\\
Inserting (\ref{FserieFm}) into (\ref{profile_qm}) after changing $l$ for $(L+1)/2+l$ gives
$${q}_m^{(\frac{L+1}{2}+l)}= \frac{q_0}{L} \sum_{s=1}^L \left[\sum_{p=0}^\infty f^{(m)}_p \cos\left(\frac{2\pi p s}{L}\right)\right] e^{\frac{2\pi\ic l s}{L}} =q_0 \sum_{p=0}^\infty f^{(m)}_p \left[
\frac{1}{L} \sum_{s=1}^L  \cos\left(\frac{2\pi p s}{L}\right) e^{\frac{2\pi\ic l s}{L}}\right]\ . $$
Using (\ref{orthocosexp}) one gets
\begin{equation}
    {q}_m^{(\frac{L+1}{2}+l)}=q_0 \frac{f_l^{(m)}}{2},\qquad l \ge 0\ .
\end{equation}
Similarly
$$ {q}_m^{(\frac{L+1}{2}-l)}=q_0 \frac{f_l^{(m)}}{2}, \qquad 0\le l \le (L-1)/2\ . $$
This can be rewritten
\begin{equation}
\hspace{-1cm}q_m^{(\frac{L+1}{2}\pm l)}= \frac{q_0}{2} f_l^{(m)}= \frac{q_0}{L} \frac{1}{1+\delta_{l,0}} \int_0^L F_m(s) \cos\left(\frac{2\pi l s}{L}\right)\ud s,\quad 0\le l\le \frac{L-1}{2} .\label{qm_int}
\end{equation}

To perform asymptotics in the regime $m,L\to\infty$ we will need another definition of (\ref{defFm}). Notice that the polynomials $P_m$ are closely related to standard orthogonal polynomials.
Chebyshev polynomials of the second kind $U_n$ are defined by the following formula
\begin{equation}
    \forall n\in\mathbb{N},\quad U_n(\cos \theta)=\frac{\sin[(n+1)\theta]}{\sin\theta}, \sin\theta\neq 0\ .\label{defUn}
\end{equation}
One can see in particular that
$$ U_0(X)=1,\ U_1(X)=2X,\ U_2(X)=4X^2-1\ .$$
They obey multiple properties which can be found in standard textbooks. In particular, they obey a recurrence relation very close to (\ref{recurrPm}). This can be used to prove that
\begin{equation}
    \forall m \ge 1, P_m(X)=U_{m-1}\left(\frac{X}{2}+1\right) \ .
\end{equation}
From this equation and the fact that $D_s^2\ge 0$ it is more convenient to rewrite (\ref{defUn}) with $\theta=\ic y$:
$$ U_n(\cosh y)=\frac{\sinh[(n+1)y]}{\sinh y}\ .$$
This gives another expression for $F_m(s)$:
\begin{equation*}
    F_m(s)= \frac{\sinh[ m \,y(s) ]}{\sinh y(s)}- \frac{\sinh[( m -1)\, y(s) ]}{\sinh y(s)}\ ,
\end{equation*}
where a new function $y(s)$ has been introduced. It is defined implicitly by
\begin{equation*}
    2\sinh\left(\frac{y(s)}{2}\right)=D_s=k+2g\cos\frac{2\pi s}{L}\ .
\end{equation*}
In particular one 
has
$$ 1+\frac{D_s^2}{2}=\cosh y(s)\ .$$

Considering (\ref{qm_int}) and the definition of $y(s)$, it is more convenient to perform the change of variable
$$ s \mapsto \theta,\ \theta=\frac{2\pi s}{L}\ .$$
such that each position along the chain is now given by the integral
\begin{equation}
\forall m\ge 1, \quad q_m^{(\frac{L+1}{2}\pm l)}= \frac{q_0}{2\pi} \frac{1}{1+\delta_{l,0}} \int_0^{2\pi} F_m(\theta) \cos\left( l \theta\right)\ud \theta,\ 0\le l\le \frac{L-1}{2} .\label{qm_int2}
\end{equation}
with now
\begin{eqnarray}
   F_m(\theta)&=& \frac{\sinh[ m \,y(\theta) ]}{\sinh [y(\theta)]}- \frac{\sinh[( m -1)\, y(\theta) ]}{\sinh [y(\theta)]}\ ,  \\ 
    2\sinh \left(\frac{y(\theta)}{2}\right)&=&k+2g\cos\theta\ .\label{def_y_theta}
\end{eqnarray}

The formula (\ref{qm_int2}) is exact and stands for another result of our study. It can be used for an asymptotic expansion for the profile. For $m\to\infty$ the following approximation is used (taking $y(\theta)\ge 0$ for simplicity)
\begin{equation}
\hspace{-2cm}F_m(\theta)=\cosh[m\, y(\theta)] -\tanh\left(\frac{y(\theta)}{2}\right) \sinh[m\, y(\theta) ]\simeq \frac{e^{m \, y(\theta)}}{2} \left[ 1- \tanh\left(\frac{y(\theta)}{2}\right)\right]
\end{equation}
Inserting it into (\ref{qm_int2}) the integral can be estimated via the saddle point method for $1\ll m$ and $1\ll l\ll L$. Start from the approximated position profile
\begin{equation*}
    q_m^{(\frac{L+1}{2}\pm l)}\simeq \frac{q_0}{2\pi} \int_0^{2\pi} \frac{e^{m \, y(\theta)}}{2} \left[ 1- \tanh\left(\frac{y(\theta)}{2}\right)\right] \cos\left( l \theta\right)\ud \theta\ .
\end{equation*}
The saddle point method requires to find the maximum of $y(\theta)$. From its implicit definition (\ref{def_y_theta}) one can deduce
$$ y'(\theta)=-2g\frac{\sin\theta}{\sqrt{ 1 +\frac{1}{4}\left( k+2g\cos\theta\right)^2 }} \ .
$$
Estimating the second derivative shows that $y(\theta)$ reaches its maximum for $\theta=0$ when $g\ge 1$ and 
$$ y''(0)=-\frac{4g}{\sqrt{ 4 +(k+2g)^2 }}\ . $$
The position profile then becomes
\begin{equation}
    q_m^{(\frac{L+1}{2}\pm l)}\simeq \frac{q_0}{2\sqrt{2\pi m |y''(0)|}}\left[1-\tanh\left(\frac{y_0}{2}\right)\right]e^{m y_0},\label{profile_approx}
\end{equation}
where $y_0=y(0)=2\,\arcsinh\left(k/2+g\right)$. In particular the position profile becomes constant, independent of the site index $l$, within this approximation.

\subsection{Diffusive transport across the chain}

The expression (\ref{profile_approx}) when, taken modulo $1$, becomes such a rapidly varying function (of $q_0$) that it can be approximated by a random variable uniformly distributed in $[0:1)$ when averaging over the initial conditions. More precisely
\begin{equation*}
   \left< q_m^{(\frac{L+1}{2}\pm l)}\right>\simeq \int_0^1 x\ud x=\frac{1}{2}\ .
\end{equation*}
This justifies the profile observed in Figs.~\ref{Ensemble Average} and \ref{Time Average}.
Using this approximation along the whole light cone yields the following estimate for the fluctuations around the averaged profile in the long time limit:
\begin{equation}
    \varepsilon_m\equiv \sum_{l=1}^L \left<\left( q_m^{(l)}-\left< q_m^{(l)}\right>\right)^2\right> \simeq 
    (l_+-l_-)\int_0^1 \left(x-\frac{1}{2}\right)^2\ud x=r\frac{m-1}{3}\ ,\label{test_diff}
\end{equation}
where the bounds of the light cone (\ref{def_light_cone_arbitrary_r}) were used.
In words we have found that the squared fluctuations around the stationary profile are linearly growing with time. This is the signature of diffusive transport along the chain with the diffusion coefficient
\begin{equation}
    D=\frac{r}{3}, \quad k\ge 1, \ g\ge 1\ .\label{def_coeff_diff}
\end{equation}
A similar computation leads to the prediction of diffusion in the momentum profile with the same diffusion coefficient.
Using this, the quantity defined in (\ref{test_diff}) can be seen as the mean distance reached at time $m$ by a trajectory in the $2L-$dimensional phase space. Its linear growth indicates a diffusive process in this high-dimensional space.

\begin{figure}[!ht]
     \begin{minipage}[left]{0.49\linewidth}
     \begin{center}
       \includegraphics[width=0.9\linewidth]{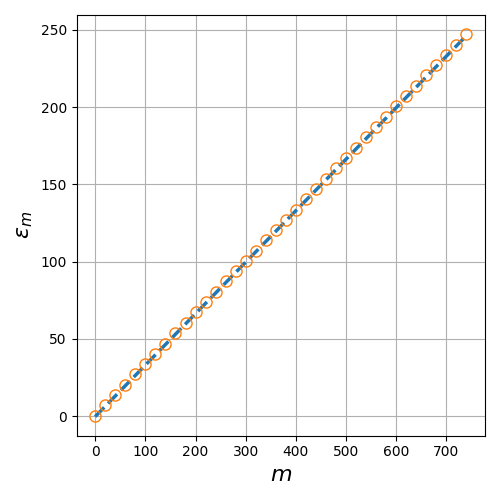}
     \end{center}
    \end{minipage}
    \begin{minipage}[right]{0.49\linewidth}
      \begin{center}
       \includegraphics[width=0.9\linewidth]{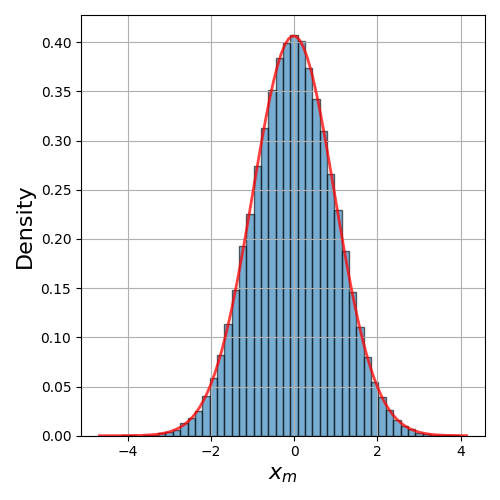}
     \end{center}
    \end{minipage}
    \caption{Left: Growth of the fluctuations of the profile for $L=3001$, $k=1$, $g=1$. The average is performed over $10^4$ initial conditions. Orange open circles stand for the numerically computed $\epsilon_m$. The dashed blue curve stands for our prediction (\ref{test_diff}). Right: Normalised histogram of the variables $x_m$ defined in (\ref{def_xm}) for the same values of $L,k,g$. The number of initial conditions is $10^6$ and $m=751$. The red curve stands for the probability density of a Gaussian random variable with zero mean and unit variance.}\label{diff}
\end{figure}

We tested our approximation numerically 
and found an excellent agreement, see Fig.~\ref{diff}. We further found that the fluctuations around the stationary profile at a fixed (long) time defined by
\begin{equation}
  x_m=\sum_{l=1}^L\left( q_m^{(l)}-\left< q_m^{(l)}\right>\right)/\sqrt{2m/3}\label{def_xm}
\end{equation}
follows a Gaussian distribution when averaging over initial conditions. It also parametrises the trajectory for the diffusive process in the phase space mentioned above.
A similar agreement is obtained for the momentum profile (data not shown).
\section{Discussion and perspectives}
\label{discuss}

We believe that the construction for coupled cat maps detailed in \cite{axenides2023arnol} is very promising. It defines a model which is strongly chaotic (uniformly hyperbolic on a compact phase space) while allowing an explicit exact expression for key quantities (Lyapunov exponents). We pursed its study for the case of a one-dimensional chain and adapted the Hopf method to prove it is ergodic. This is compatible with the stronger claim made in \cite{axenides_exponential_2024}. 

This model provides another example of a local many-body classical chaotic system. It may be used to test the foundations of statistical physics, e.g. to see whether Stefan-Boltzmann law can be explained using only classical arguments as in \cite{wang_classical_2022}. An hydrodynamic approach to describe the long time/large scale properties is another possible route to explore.
It may be also interesting to see whether the transport problem as defined here can be similarly treated for the model of coupled cat maps introduced in \cite{gutkin2016classical}. The method described in \cite{gutkin_linear_2021} to build a symbolic dynamics may also be applicable for our choice of the coupling. 
\ack

RD would like to thank Prof.~Herbert Spohn for providing the reference \cite{Dorfman}. He also benefitted from stimulating discussions with Dr~Takato Yoshimura and Prof.~Jon Keating.

\bigskip
\bibliographystyle{vancouver}
\bibliography{bibfile}

\begin{thebibliography}{10}

\bibitem{landau2013statistical}
Landau LD, Lifshitz EM.
\newblock Statistical Physics: Volume 5. vol.~5.
\newblock Elsevier; 2013.

\bibitem{gaspard_diffusion_1992}
Gaspard P.
\newblock Diffusion, effusion, and chaotic scattering: An exactly solvable
  Liouvillian dynamics.
\newblock Journal of Statistical Physics. 1992;68(5):673-747.

\bibitem{Dorfman}
Dorfman JR.
\newblock An introduction to chaos in nonequilibrium statistical mechanics.
\newblock Cambridge Univ. Press; 1999.

\bibitem{de2012largest}
de~Wijn AS, Hess B, Fine B.
\newblock Largest Lyapunov exponents for lattices of interacting classical
  spins.
\newblock Physical Review Letters. 2012;109(3):034101.

\bibitem{PhysRevX.10.041017}
Pandey M, Claeys PW, Campbell DK, Polkovnikov A, Sels D.
\newblock Adiabatic Eigenstate Deformations as a Sensitive Probe for Quantum
  Chaos.
\newblock Phys Rev X. 2020 Oct;10:041017.

\bibitem{OTOC_Larkin}
Larkin AI, Ovchinnikov.
\newblock Largest lyapunov exponents for lattices of interacting classical
  spins.
\newblock Sov JETP. 1969;28:1200.

\bibitem{garcia-mata_out--time-order_2023}
{Garc}\'ia{-}Mata I, Jalabert R, Wisniacki D.
\newblock Out-of-time-order correlations and quantum chaos.
\newblock Scholarpedia. 2023;18(4):55237.
\newblock Available from:
  \url{http://www.scholarpedia.org/article/Out-of-time-order_correlations_and_quantum_chaos}.

\bibitem{susskind2004introduction}
Susskind L, Lindesay J.
\newblock Introduction To Black Holes, Information And The String Theory
  Revolution, An: The Holographic Universe.
\newblock World Scientific; 2004.

\bibitem{sekino2008fast}
Sekino Y, Susskind L.
\newblock Fast scramblers.
\newblock Journal of High Energy Physics. 2008;2008(10):065.

\bibitem{gutkin2016classical}
Gutkin B, Osipov V.
\newblock Classical foundations of many-particle quantum chaos.
\newblock Nonlinearity. 2016;29(2):325.

\bibitem{pethel2006symbolic}
Pethel SD, Corron NJ, Bollt E.
\newblock Symbolic dynamics of coupled map lattices.
\newblock Physical Review Letters. 2006;96(3):034105.

\bibitem{axenides2023arnol}
Axenides M, Floratos E, Nicolis S.
\newblock Arnol'd cat map lattices.
\newblock Physical Review E. 2023;107(6):064206.

\bibitem{ArnoldAvez}
Arnold VI, Avez A.
\newblock Ergodic problems of Classical Mechanics.
\newblock Princeton Univ. Press; 1968.

\bibitem{percival_arithmetical_1987}
Percival I, Vivaldi F.
\newblock Arithmetical properties of strongly chaotic motions.
\newblock Physica D: Nonlinear Phenomena. 1987;25(1):105-30.

\bibitem{keating_asymptotic_1991}
Keating JP.
\newblock Asymptotic properties of the periodic orbits of the cat maps.
\newblock Nonlinearity. 1991;4(2):277-307.

\bibitem{Mehta}
Mehta ML.
\newblock Matrix Theory, Selected Topics and Useful Results.
\newblock Hindusan Publ. Corp./Les \'Editions de Physique (in English); 1977.

\bibitem{RevModPhys.57.617}
Eckmann JP, Ruelle D.
\newblock Ergodic theory of chaos and strange attractors.
\newblock Rev Mod Phys. 1985 Jul;57:617-56.
\newblock Available from:
  \url{https://link.aps.org/doi/10.1103/RevModPhys.57.617}.

\bibitem{axenides_exponential_2024}
Axenides M, Floratos E, Nicolis S. Exponential mixing of all orders for Arnol'd
  cat map lattices. {arXiv}; 2024.
\newblock Available from: \url{https://arxiv.org/abs/2401.08521}.

\bibitem{wang_classical_2022}
Wang J, Casati G, Benenti G.
\newblock Classical Physics and Blackbody Radiation.
\newblock Physical Review Letters. 2022;128(13):134101.

\bibitem{gutkin_linear_2021}
Gutkin B, Cvitanović P, Jafari R, Saremi AK, Han L.
\newblock Linear encoding of the spatiotemporal cat.
\newblock Nonlinearity. 2021;34(5):2800-36.

\end{thebibliography}

\appendix
\section{Computational Methods}
\label{comput_methods}
In the field of dynamical systems, numerical methods are indispensable not only as a means to visualize, but as a tool for validating and testing theoretical models. However, the limitation of computational precision, particularly in floating-point arithmetic, poses a significant challenge, as small errors can exponentially amplify, leading to divergent system behaviour. Mitigating this is crucial for maintaining fidelity in the representation of the system, unfortunately often at the cost of increased computational demand.

\subsection{General Approach}
By sectioning up the parameter space we can craft individual numerical methods that are significantly more efficient at testing the system within their domain, this also adds a level of validation to the models as we can test models with overlapping parameter spaces against each other.

\subsection{Python Implementation}

The bulk of our calculations was conducted using Python. While the code is logically structured, some sections deviate from best practices, affecting only the efficiency and readability but not the accuracy of the output. The below are the functions used to calculate a "dataframe" that contains the evolution of the system based off the parameters of the system. The code for our simulations is available upon request.

\subsubsection{Generalised Matrix Model}

We begin with a generalized model for simulating the system, constructing matrix \( M \) as outlined in \cite{axenides2023arnol}. The use of the `mpmath` package allows for high-precision calculations to handle the chaotic nature of the system. This model is valid for the majority of parameter space but is very computationally expensive. Post-simulation, data is converted back to double precision to manage computational resources.

\subsubsection{Vectorised Model}

For the case of nearest neighbour coupling we can simplify the system into a set of vectors, the same approach could be taken for all coupling ranges but produces diminishing improvements in efficiency for increasing complexity of equations.

\subsubsection{Keeping rational numbers and doing integer computations}
The two above methods make use of mpmath package which allows for arbitrary precision which is vital when sampling from the real numbers, however it is very expensive computationally.
Focusing on rational initial conditions, we leverage Python's capability for arbitrary-precision integers, significantly enhancing computational speed. While this method limits sampling from the reals, it allows us to sample from various rational numbers


Statistical calculations, both ensemble and time averages, are integrated into our computational methods, ensuring consistency across different approaches.

\subsection{Performance Metrics}

\begin{figure}[!ht]
     \begin{minipage}[left]{0.49\linewidth}
      \begin{center}
        \includegraphics[width=0.9\linewidth]{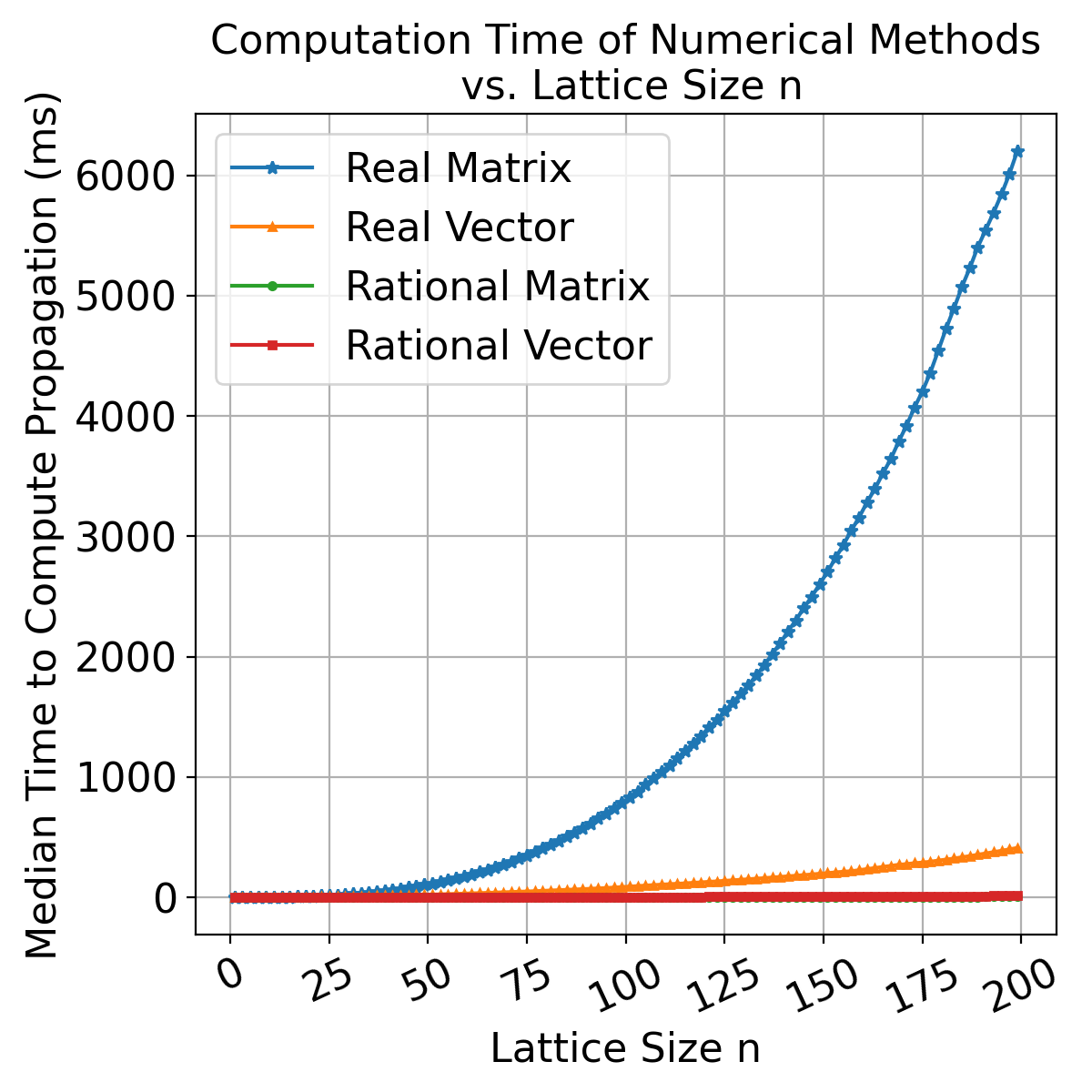}
      \end{center}
    \end{minipage}
    \begin{minipage}[right]{0.49\linewidth}
      \begin{center}
        \includegraphics[width=0.9\linewidth]{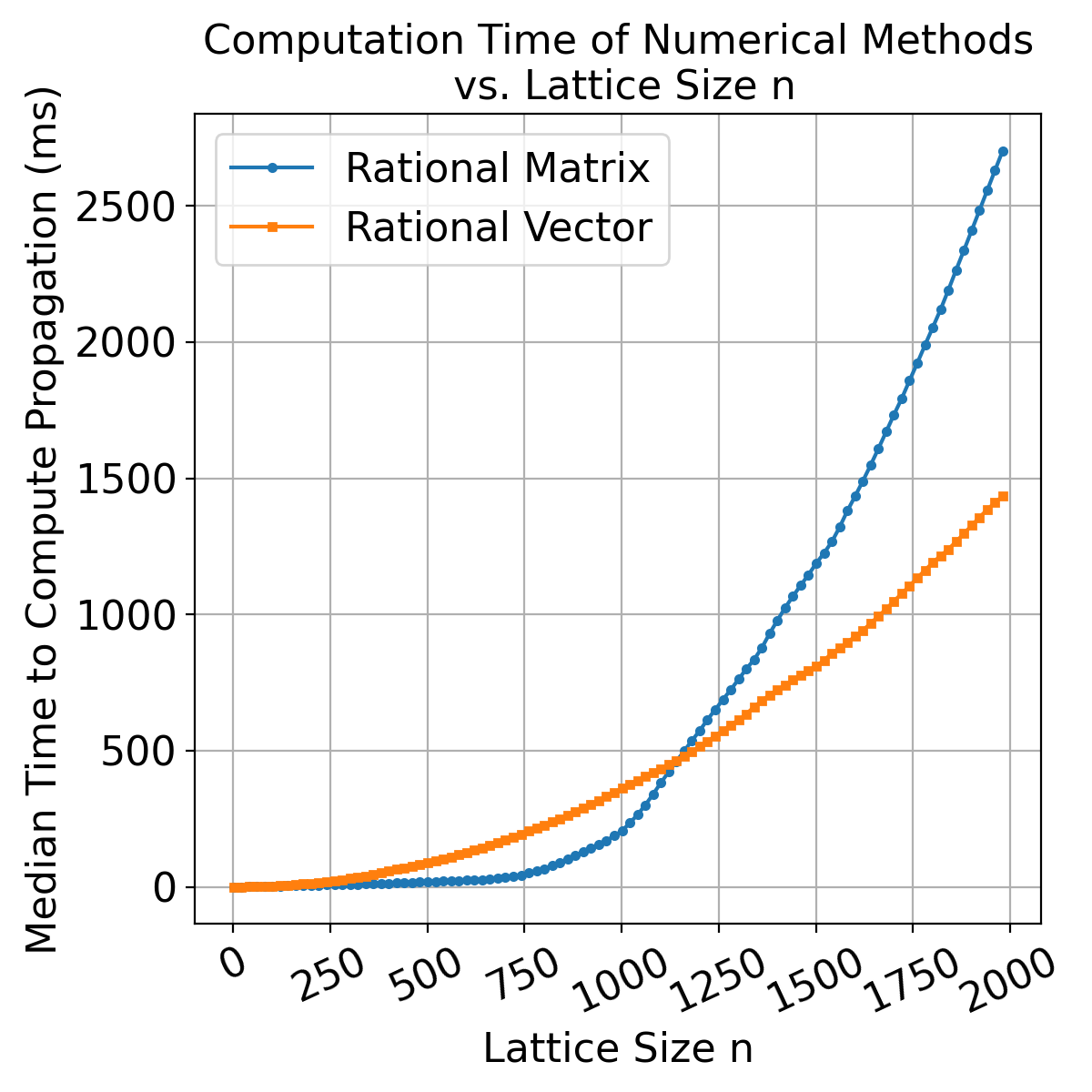}
      \end{center}
    \end{minipage}
    \caption{The time taken to generate the numerical values of a central perturbation propagating to the edge of a chain of length n for the 4 numerical methods, averaged across 1000 I.C.s}
\end{figure}

As can be seen above there is a large difference in the efficiency of different methods, however there are times when the less efficient methods are the only ones covering the required parameter space.


\section{Exact solution for the Initial Value Problem}
\label{IVP}

Here are repeated the steps to solve the initial value problem. They are given for the sake of self consistency and using the notation in the main text. In particular each point in the phase space belonging to a trajectory at time $m$ is identified with the $2L-$dimensional vector $\vec{X}_m$, which following (\ref{defX}), shall be written in a more compact form as
\begin{equation}
  \vec{X}_m\left(\begin{array}{c} \vec{q}_m \\ \vec{p}_m\end{array}\right) \ .
\end{equation}
The dynamical equations for the phase space variables are following (\ref{defFibchaincat}), with $m\ge 0$,
\begin{eqnarray}
  \vec{q}_{m+1}=&\vec{q}_m+C\vec{p}_m\quad &\textrm{mod } 1, \label{eq_qm}\\
  \vec{p}_{m+1}=& C\vec{q}_m+(\one_L+C^2)\vec{ p}_m\quad  &\textrm{mod } 1.\label{eq_pm}
\end{eqnarray}
Solving for $\vec{p}_m$ in (\ref{eq_qm})
$$ \vec{p}_m=C^{-1}(\vec{q}_{m+1}-\vec{q}_{m}) $$
then inserting it into (\ref{eq_pm}) leads to
$$ C^{-1}(\vec{q}_{m+2}-\vec{q}_{m+1})=C\vec{q}_m+(1+C^2)C^{-1}(\vec{q}_{m+1}-\vec{q}_{m}) . $$
Left multiply by C, and using $C^{-1}C^2=C^2C^{-1}$,
$$ (\vec{q}_{m+2}-\vec{q}_{m+1})=C^2\vec{q}_m+(1+C^2)(\vec{q}_{m+1}-\vec{q}_{m}) $$
which is after rearranging
\begin{equation}
  \vec{q}_{m+2}=(2+C^2)\vec{q}_{m+1}-\vec{q}_m,\quad m\ge 0\ . \label{Newtonlaw_qm}
\end{equation}
This equation is equivalent to Newton's second law for this discrete map.
The same difference equation can be derived in a similar manner for the momentum $\vec{p}_m$.

Due to the choice of periodic boundary conditions on the chain, those difference equations can be solved via discrete Fourier transform. It comes from the fact that $C$ is the sum of shift matrices, hence becomes diagonal in the Fourier basis. 
More precisely, if one introduces the discrete Fourier transform matrix of size $L\times L$
 \begin{equation}
     F=\left(F_{lm}\right), \quad F_{lm}=\frac{e^{2\pi \ic l m/L}}{\sqrt{L}}\Rightarrow F F^\dag=\one_{L}\ . \label{defmatF}
 \end{equation}
 The matrix $F$ contains the eigenvectors of $C$ and one has
 \begin{equation*}
     F^\dag CF= D=\left(\begin{array}{cccc}
    k+ 2g\cos\left(\frac{2\pi}{L}\right)&0&\dots &0 \\
    0& k+ 2g\cos\left(\frac{4\pi }{L}\right) &\dots &0 \\
    \vdots&\vdots&\ddots&\vdots\\
    0&0&\dots &k+ 2g\cos\left(\frac{2\pi L }{L}\right)
    \end{array}\right)
 \end{equation*}
 For convenience, the eigenvalues of $C$ shall be denoted as
 \begin{equation*}
     D_l=k+2g \cos\left(\frac{2\pi l}{L}\right),\quad 1\le l \le L \ .
\end{equation*}
The change of basis enables one to solve (\ref{Newtonlaw_qm}). This is proven as follows.
To avoid the presence of zero modes, one shall now only take {$L$  an odd integer}.
Change the variable for $\vec{r}_m=F^\dag \vec{q}_m$ in (\ref{Newtonlaw_qm}). This latter equation becomes
$$F\vec{r}_{m+2}-2F\vec{r}_{m+1}+F\vec{r}_{m}=C^2F\vec{r}_{m+1} \ .$$
Left multiply by $F^\dag$, and use the unitarity of $F$ to get
$$\vec{r}_{m+2}-2\;\vec{r}_{m+1}+\vec{r}_{m}=F^\dag C^2F\vec{r}_{m+1}\ .$$
Since $F$ diagonalizes $C$, it does so for any of its powers
$$ D=F^\dag C F \Rightarrow F^\dag C^t F=D^t\ \forall t\in \mathbb{N} .$$    
The equation of dynamics for the new variables simplifies to
$$\vec{r}_{m+2}-2\;\vec{r}_{m+1}+\vec{r}_{m}=D^2\;\vec{ r}_{m+1}\ .$$
The benefit of the change of variables becomes clear: the system of difference equations becomes diagonal and one can solve for each coordinate independently. For the $l$th coordinate $r_{m}^{(l)}$ of $\vec{r}_m$, the equation to solve is
\begin{equation}
    r_{m+2}^{(l)}-2\; r_{m+1}^{(l)}+ r_{m}^{(l)}=D_l^2\;r_{m+1}^{(l)}\ .\label{eq_rm}
\end{equation}
Trying the ansatz
\[ r_{m}^{(l)}=a_l \rho_l^m\]
leads to
\[a_l \rho_l^m\left[ \rho_l^{2}-\left(2+D_l^2\right) \rho_l+1\right]=0\ . \]
In order to have a nontrivial solution ($a_l\neq 0,\ \rho_l\neq 0$), one must require
    \[\rho_l^{2}-(2+D^2_l)\rho_l+1=0 \ .\]
whose solutions are directly given by
$$\lambda_{l}=\frac{2+D_l^2}{2}+|D_l|\frac{\sqrt{D^2_l+4}}{2}, \frac{1}{\lambda_l} \ .$$
As the difference equation (\ref{eq_rm}) is linear, one can use the superposition principle to write its general solution
\begin{equation*}
    {r}_{m}^{(l)}=a_{+}^{(l)} \lambda_{l}^m+a_{-}^{(l)}\lambda_{l}^{-m},\quad m\ge 0,
\end{equation*}
from which the full solution $\vec{r}_m$ is deduced in a matrix form
 \[\left(\begin{array}{c}
    r_{m}^{(1)} \\
    r_{m}^{(2)} \\
    \vdots \\
    r_{m}^{(L)}
    \end{array}\right)=\left(\begin{array}{cccc}
    \lambda_{1}^m &0&\dots &0 \\
    0&\lambda_{2}^m&\dots &0 \\
    \vdots&\vdots&\ddots&\vdots\\
    0&0&\dots &\lambda_{L}^m
    \end{array}\right)\left(\begin{array}{c}
    a_+^{(1)} \\
    a_+^{(2)} \\
    \vdots \\
    a_+^{(L)}
    \end{array}\right) +\left(\begin{array}{cccc}
    \lambda_{1}^{-m} &0&\dots &0 \\
    0&\lambda_{2}^{-m}&\dots &0 \\
    \vdots&\vdots&\ddots&\vdots\\
    0&0&\dots &\lambda_{L}^{-m}
    \end{array}\right)\left(\begin{array}{c}
    a_-^{(1)} \\
    a_-^{(2)} \\
    \vdots \\
    a_-^{(L)}
    \end{array}\right)
   \ . \]
Or, more concisely,
\[ \vec{r}_m = \Lambda^m\; \vec{a}_+ +  \Lambda^{-m}\; \vec{a}_-, \]
where the $L\times L$ diagonal matrix $\Lambda$ has been introduced which contains the $\lambda_l$ along the diagonal.
To solve the general solution problem, one needs to expand the vectors $\vec{a}_\pm$ as functions of the initial state $\vec{q}_0$ and $\vec{p}_0$. This procedure has to be done separately for $\vec{q}_m$ and $\vec{p}_m$. 

After going back to the initial variables via inverse Fourier transform one gets
\begin{eqnarray}
    \vec{q}_m =& F\left[ \left(\Lambda^{-1} -\Lambda \right)^{-1} \left( \Lambda^{m-1} - \Lambda^{-(m-1)} + \Lambda^{-m} - \Lambda^m \right) \right] F^\dag \vec{q}_0&\nonumber \\
    & + F\left[ \left(\Lambda^{-1} -\Lambda \right)^{-1} D\left( \Lambda^{-m} - \Lambda^m \right) \right] F^\dag \vec{p}_0 \quad \textrm{  mod }1 \label{IVP_q}\ ,\\
   \vec{p}_m =& F\left[ \left(\Lambda^{-1} -\Lambda \right)^{-1} \left( \Lambda^{m-1} - \Lambda^{-(m-1)} + (1+D^2)(\Lambda^{-m} - \Lambda^m )\right) \right] F^\dag \vec{p}_0&\nonumber \\
    & +F\left[ \left(\Lambda^{-1} -\Lambda \right)^{-1} D\left( \Lambda^{-m} - \Lambda^m \right) \right] F^\dag \vec{q}_0\quad  \textrm{  mod }1\label{IVP_p} \ .
\end{eqnarray}

\section{Perturbed cat map}
\label{perturbed}

\begin{figure}[H]
\centering
\begin{subfigure}{\subfigwidth}
  \centering
  \includegraphics[width=1\linewidth]{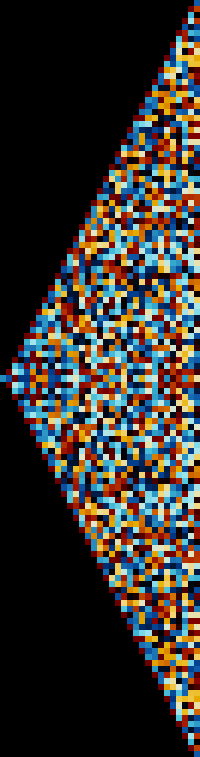}
  \caption{}
\end{subfigure}
\begin{subfigure}{\subfigwidth}
  \centering
  \includegraphics[width=1\linewidth]{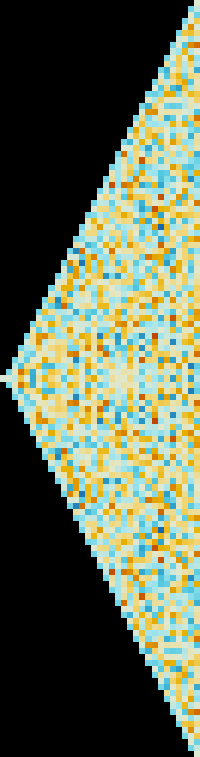}
  \caption{}
\end{subfigure}
\begin{subfigure}{\subfigwidth}
  \centering
  \includegraphics[width=1\linewidth]{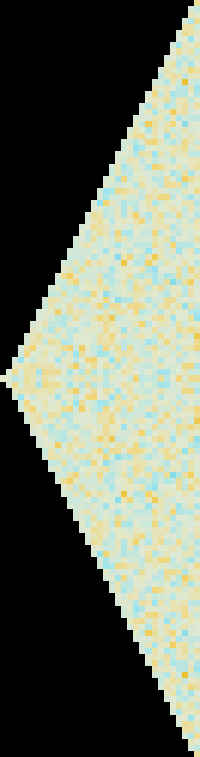}
  \caption{}
\end{subfigure}
\begin{subfigure}{\subfigwidth}
  \centering
  \includegraphics[width=1\linewidth]{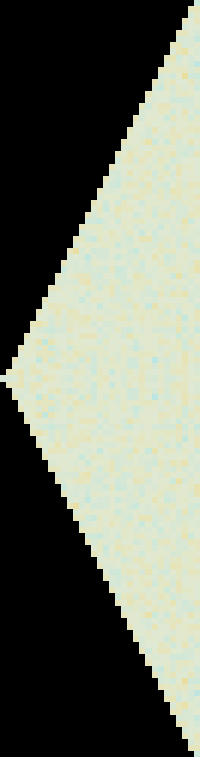}
  \caption{}
\end{subfigure}
\begin{subfigure}{\subfigwidth}
  \centering
  \includegraphics[width=1\linewidth]{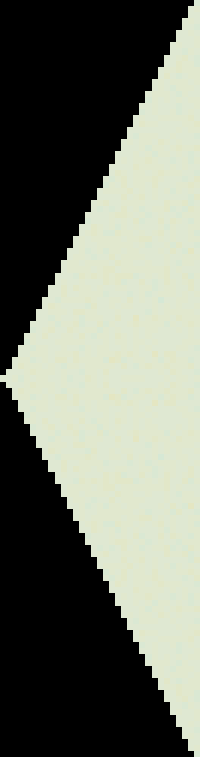}
  \caption{}
\end{subfigure}
\caption{Space-time diagrams for the perturbed mapping (\ref{def_perturbed_cat_map}) with the initial condition (\ref{IC_chain}) for $k=1$, $g=1$, $L=125$, $\epsilon=0.1$ across various levels of sampling, illustrating convergence towards the equilibrium state for Samples $=1, 10, 100, 1000, 10000$ going from left (a) to right (e) respectively.}
\label{Perturbed Ensemble Average}
\end{figure}

\begin{figure}[H]
\centering
\begin{subfigure}{\subfigwidth}
  \centering
  \includegraphics[width=1\linewidth]{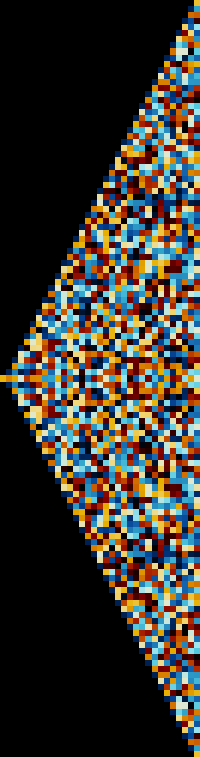}
  \caption{}
\end{subfigure}
\begin{subfigure}{\subfigwidth}
  \centering
  \includegraphics[width=1\linewidth]{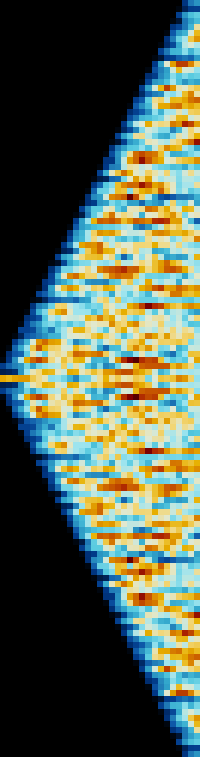}
  \caption{}
\end{subfigure}
\begin{subfigure}{\subfigwidth}
  \centering
  \includegraphics[width=1\linewidth]{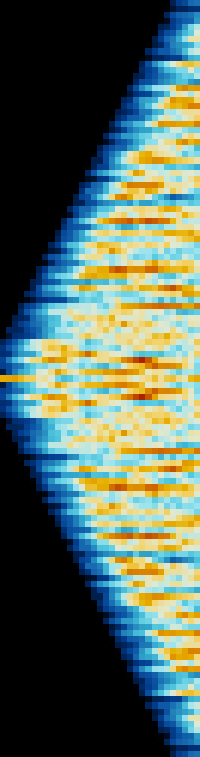}
  \caption{}
\end{subfigure}
\begin{subfigure}{\subfigwidth}
  \centering
  \includegraphics[width=1\linewidth]{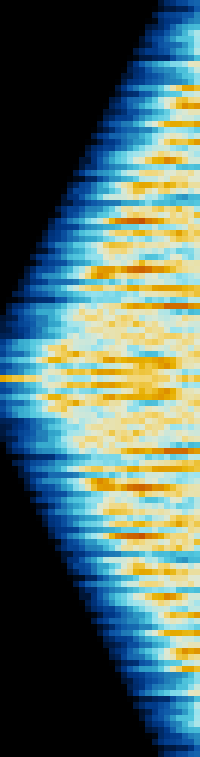}
  \caption{}
\end{subfigure}
\begin{subfigure}{\subfigwidth}
  \centering
  \includegraphics[width=1\linewidth]{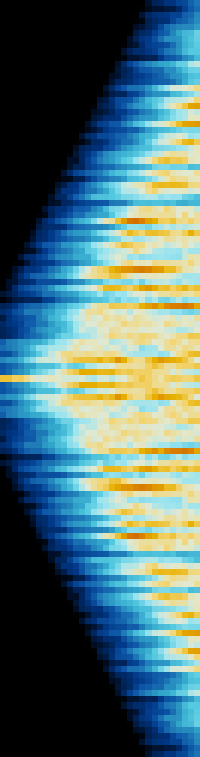}
  \caption{}
\end{subfigure}
\caption{Space-time diagrams for the perturbed mapping (\ref{def_perturbed_cat_map}) with the initial condition (\ref{IC_chain}) for $k=1$, $g=1$, $L=125$, $\epsilon=0.1$ across various levels of time averaging, illustrating convergence towards the equilibrium state for $\Delta m=1, 5, 9, 13, 17$ going from left (a) to right (e) respectively.}
\label{Perturbed Time Average}
\end{figure}

It is well known that the single cat map for $L=1$ can have highly degenerate periodic orbits due to the number-theoretic properties. In order to remove this effect, it is customary to add a small perturbation for which the map retains its chaotic properties. First the perturbed cat map for one cell is defined by 
\begin{eqnarray*}
    q_{m+1} &=& q_m+ k p_m - 
    \epsilon \sin p_{m+1}\\
    p_{m+1}&=& k q_m+ (k^2+1) p_m
\end{eqnarray*}
One can generalize it in the coupled case (for nearest neighbour coupling, build similarly as in the main text) via the following nonlinear map:
\begin{equation}
    \vec{X}_{m+1}=\vec{F_\epsilon}(\vec{X}_m) \label{def_perturbed_cat_map}
\end{equation}
which reads in detail
\begin{eqnarray}
\hspace{-2cm}q_{m+1}^{(l)}=&q_m^{(l)}+ k p_m^{(l)} - \epsilon \sin\left[2\pi p_{m+1}^{(l)}\right]+ g\left[ p_m^{(l-1)}+p_m^{(l+1)}\right]
\label{perturb_eq_qm_full}\\
\hspace{-2cm}p_{m+1}^{(l)}=&  k q_m^{(l)}+ g\left[ q_m^{(l-1)}+q_m^{(l+1)}\right] +(k^2+1+2g^2) p_m^{(l)}+  2kg\left[ p_m^{(l-1)}+p_m^{(l+1)}\right]+g^2\left[ p_m^{(l-2)}+p_m^{(l+2)}\right]&
\label{perturb_eq_pm_full},
\end{eqnarray}
where the mod $1$ operation is implicit for each equation.
Figure~\ref{Perturbed Ensemble Average} shows that the same profile occurs when averaging over initial conditions than for the unperturbed map.
Similarly Figure~\ref{Perturbed Time Average} shows that the same profile occurs when averaging over a time window than for the unperturbed map.

\end{document}